\documentclass[12pt]{iopart}

\expandafter\let\csname equation*\endcsname\relax
\expandafter\let\csname endequation*\endcsname\relax

\usepackage{amsmath,amssymb}
\usepackage[dvipsnames,svgnames,x11names]{xcolor}
\usepackage{graphicx}                
\usepackage{subfigure}
\usepackage{siunitx}
\usepackage{array}

\usepackage[T1]{fontenc}
\usepackage[utf8]{inputenc}

\usepackage[numbers]{natbib}

\newcommand{\im}{\mathrm{i}}
\newcommand{\ce}{\mathrm{e}}

\newcommand{\pd}[2]{\frac{\partial #1}{\partial #2}}
\newcommand{\pdd}[2]{\frac{\partial^2 #1}{\partial #2^2}}

\newcommand{\FT}{_{\textrm{FT}}}
\newcommand{\Up}{U_{\textrm{p}}}
\newcommand{\Ip}{I_{\textrm{p}}}
\renewcommand{\vec}[1]{\mathbf{#1}}
\newcommand{\Ordo}{\mathcal{O}}

\newcommand{\intensity}[1]{\SI{#1}{\watt\per\centi\meter\squared}}

\usepackage[colorlinks=true, linkcolor=FireBrick, urlcolor=FireBrick, citecolor=ForestGreen]{hyperref}

\begin{document}

\title[Spatially and spectrally resolved quantum path interference\dots]{Spatially and spectrally resolved quantum
  path interference with chirped driving pulses}

\author{Stefanos~Carlstr\"{o}m${^{1,6,*}}$,
  Jana~Precl\'{\i}kov{\'a}$^{1,2,6}$,
  Eleonora~Lorek$^1$,
  Esben~Witting~Larsen$^1$,
  Christoph~M~Heyl${^1}$,
  David~Pale\v{c}ek${^{3,4}}$,
  Donatas~Zigmantas{$^3$},
  Kenneth~J~Schafer{$^5$},
  Mette~B~Gaarde{$^5$},
  and~Johan~Mauritsson$^{1,\dagger}$}

\address{$^1$Department of Physics, Lund University, Box 118, 222 10
  Lund, Sweden}
\address{$^2$Department of Physics and Technology, University of
  Bergen, 5007 Bergen, Norway}
\address{$^3$Department of Chemical Physics, Lund University, Box 124,
  222 10 Lund, Sweden}
\address{$^4$Department of Chemical Physics, Charles University in
  Prague, Ke~Karlovu~3, 121~16~Praha~2, Czech Republic.}
\address{$^5$Louisiana State University, Baton Rouge, 70803-4001,
  Louisiana, United States of America.}
\address{$^6$These authors contributed equally to this work.}

\eads{$^*$\mailto{stefanos.carlstrom@fysik.lth.se},
  $^\dagger$\mailto{johan.mauritsson@fysik.lth.se}}

\begin{abstract}
  We measure spectrally and spatially resolved high-order harmonics
  generated in argon using chirped multi-cycle laser pulses. Using a
  stable, high-repetition rate laser we observe detailed interference
  structures in the far-field. The structures are of two kinds;
  off-axis interference from the long trajectory only and on-axis
  interference including the short and long trajectories. The former
  is readily visible in the far-field spectrum, modulating both the
  spectral and spatial profile. To access the latter, we vary the
  chirp of the fundamental, imparting different phases on the
  different trajectories, thereby changing their relative phase. Using
  this method together with an analytical model, we are able to
  explain the on-axis behaviour and access the dipole phase parameters
  for the short (\(\alpha_s\)) and long (\(\alpha_l\)) trajectories. The
  extracted results compare very well with phase parameters calculated
  by solving the time-dependent Schrödinger equation. Going beyond the
  analytical model, we are also able to successfully reproduce the
  off-axis interference structure.
\end{abstract}

\vspace{2pc}

\submitto{\NJP}
\maketitle

\section{Introduction}
High-order harmonic generation (HHG) is a nonlinear optical process,
in which a comb consisting of multiples of the driving laser frequency
\(\omega_0\) is emitted coherently after interaction with a
target~\cite{McPherson1987,Ferray1988}. HHG and the understanding of
the process itself has led to the field of attosecond
physics~\cite{Corkum2007}, which enables the time-resolved observation
of electron
dynamics~\cite{Drescher2002,Mauritsson2010PRL,Schultze2010S,Kluender2011}.

The HHG process can be understood using a semi-classical three step
model in which an electron is first ionized by tunnelling, is
subsequently accelerated in the laser field, and finally returns to
the ion core and upon recombination releases its excess kinetic energy
leading to the emission of high energy photons
\cite{Schafer1993,Corkum1993}. The generated harmonics are of odd
orders since the process is repeated every half cycle of the laser
field. This semi-classical understanding has been verified extensively
through comparison with experiments and with more sophisticated
calculations based on the integration of the time-dependent
Schrödinger equation (TDSE) within the single-active-electron (SAE)
approximation, either in its full numerical
form~\cite{KrauseSchaferKulander1992} or within the strong field
approximation (SFA)~\cite{Lewenstein1994}. From this three-step model
for HHG, it follows that for each harmonic energy there are multiple
quantum paths (QPs) the electron can follow in the continuum. They
correspond to different pairs of ionization and return times ($t_i$,
$t_r$), that give rise to the same kinetic energy upon return. The two
first QPs, termed the short and long QPs, both return within one cycle
after ionization, with the short QP being released later and returning
earlier than the long QP. The emission generated from each of these
two QP contributions has different macroscopic coherence
properties~\cite{Bellini1998,Lyngaa1999,Gaarde1999} because of the
different microscopic phase that is imparted via the semi-classical
action accumulated along each path. As we will describe in more detail
below, this phase is approximately proportional to the cycle-averaged
laser intensity with a phase coefficient $\alpha$ that increases with the
time spent in the continuum. This means that the intensity dependence
of the short-path contribution to the harmonic emission is much
smaller than that of the long-path contribution. Therefore, the short
QP emission has a smaller spectral and spatial divergence imparted by
the temporal and radial variation of the laser intensity in the
generation region.

The dipole properties of the HHG process may lead to various
interference effects, since the same final energy is generated from
several different trajectories. Interferences appearing as spectral
and/or spatial structures in the harmonic far-field emission have been
reported and identified as interferences between the short and the
long quantum path contribution, known as quantum path interferences
(QPI)~\cite{Benedetti2006,Corsi2006,Zair2008,Ganeev2011,He2015a}. Other
works identify interferences arising within the long path contribution
only due to spatio-temporal phase and amplitude modulations in the
generation medium~\cite{Heyl2011,Dubrouil2014} or due to the spectral
interference of adjacent harmonics~\cite{Sansone2005}. The emission
from the short QP contribution has been characterized in much more
detail~\cite{Paul2001,Mauritsson2004,Varju2005JoMO} than that from the
long QP contribution~\cite{Sansone2005} as the latter is more
difficult to accurately phase match and control
experimentally~\cite{Benedetti2006,He2015a}.

In this paper we present a detailed experimental characterization of
the phase properties of the short and long QP contributions to HHG,
via QPI in both the spectral and spatial domain which we control
through the chirp of the generating laser pulse, also changing the
pulse duration and the intensity. We identify and distinguish quantum
path interferences from macroscopic interference effects arising
within the emission of the long trajectory. We use a commercial
turn-key laser system that produces long multi-cycle pulses (with
durations \(\geq\SI{170}{\femto\second}\), corresponding to
\(\geq\)~50~cycles at the driving wavelength
\(\lambda=\SI{1030}{\nano\meter}\)), which yield spatially and spectrally
well-resolved harmonics with high signal-to-noise ratio. The stable
operation of the laser in combination with the long and controllable
pulses allow us to observe and characterize the QPI for a range of
harmonics in argon spanning from harmonic 11, which is below the
ionization threshold, to harmonic 37. In a single spectrum, clear
spatial and spectral modulation of the harmonic order is visible,
predominantly for the contribution of larger spatial and spectral
divergence, i.e.\ the long QP contribution. The interference between
the short and the long QPs, however, is not visible from one spectrum
alone, but it is sensitive to intensity variation of the driving
field. Therefore, controlling the shape of the driving pulse by adding
a frequency chirp, the HHG process is affected through the increase in
pulse duration and a decrease of the peak intensity. This reveals the
interference between the short and long QPs, since their respective
phases depend differently on the peak intensity.  Additionally, the
sign of the driving pulse chirp changes the spectral phase of the
harmonic emission and in particular influences the QPs differently.

We implement a model based on the semi-classical description of
HHG~\cite{Gaarde1999} as driven by a laser pulse, which is Gaussian in
the temporal and spatial domain. This simple model captures the
observed on-axis short--long QPI features very well and can be used to
extract the phase coefficients $\alpha_s$ and $\alpha_l$ from the experimental
results. We also compare the experimental results to numerical
calculations performed both within the SAE-TDSE and the SFA. We
measure experimental values for $\alpha_l$ in good agreement with those
obtained in~\cite{Benedetti2006,He2015}. For $\alpha_s$ we measure values
that are small and negative for a range of low-order harmonics,
indicating that the interaction between the returning electron and the
ionic potential plays a substantial role in the generation of these
harmonics. Negative values for $\alpha_s$ have been predicted in some
calculations~\cite{Mauritsson2004,Varju2005JoMO} but have not to our
knowledge been observed experimentally to this date.

We further investigate the off-axis interference structures by
employing a more complete, but numerical model. The spectral--spatial
modulation due to the long QP is very well reproduced by this model
and explains the significance of the contributions that go beyond the
analytical model, namely phase curvature effects of higher order than
parabolic and the intensity dependence of the dipole phase parameters
\(\alpha_{s/l}\).

The paper is organized as follows: in section~\ref{sec:exp-method},
the experimental method used to obtain the data is briefly outlined,
and in section~\ref{sec:exp-results}, these data are presented. In
section~\ref{sec:math-model}, the mathematical model is described and
in section~\ref{sec:tdse} the quantum mechanical calculations used to
verify our modelling are presented. Whereas
sections~\ref{sec:math-model}--\ref{sec:tdse} are mainly concerned with
the short--long QPI, section~\ref{sec:long--long} describes the
interference structures visible off-axis, where no short QP is
present. Finally, in section~\ref{sec:discussion} we discuss our
results and what we can extract from them.


\section{Experimental method}
\label{sec:exp-method}
\begin{figure}
  \begin{center}
    \includegraphics[width=\linewidth]{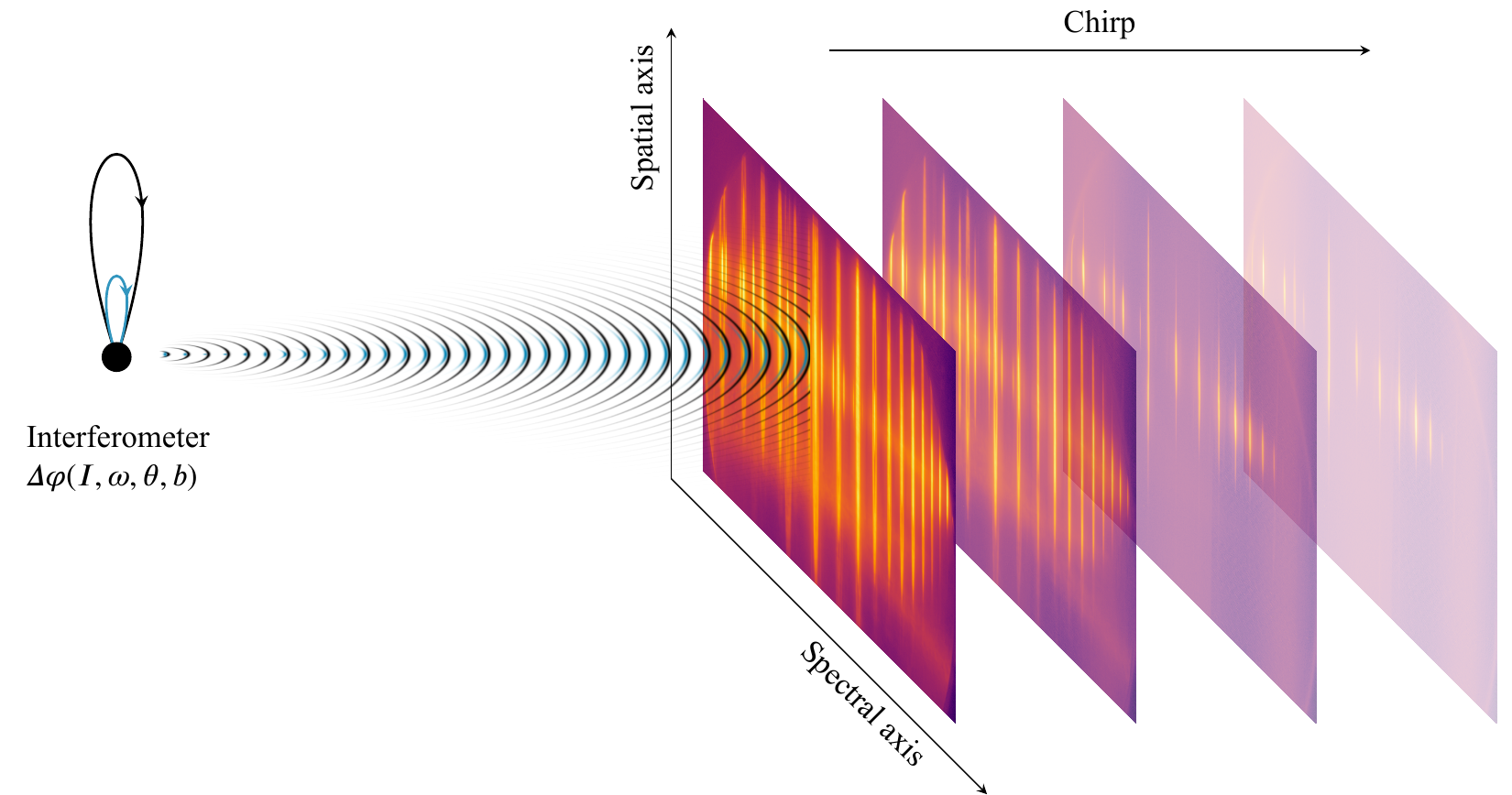}
  \end{center}
  \caption{\label{fig:Scheme} Experimental scheme; in our experiment
    the HH emission consists of contributions related to the two
    shortest QPs: short (blue) and long (black). We gradually vary
    the chirp of the driving pulses and observe spatially and
    spectrally resolved HH. The experiment can be understood as an
    interferometer, where the phase difference $\Delta \varphi$
    between the long and short trajectories varies with the driving
    intensity $I$, the emitted frequency $\omega$ and the angle of
    emission $\theta$ and the chirp parameter $b$.}
\end{figure}

The high-order harmonics (HHs) were generated in argon by a
commercially available compact Yb:KGW PHAROS laser (Light
Conversion). The pulse energy was \(\SI{150}{\micro\joule}\), the
central wavelength \(\lambda=\SI{1030}{\nano\meter}\) and the repetition
rate was set to \(\SI{20}{\kilo\hertz}\).  The pulse-to-pulse
stability of this laser is \(<\SI{0.5}{\%}\) rms over \SI{24}{hours}.
The duration and chirp of the pulses were varied by adjusting the
grating compressor. The adjustment of the grating enabled a gradual
change of the pulse duration from negatively chirped pulses of
\(\SI{500}{\femto\second}\) to Fourier-transform (FT) limited pulses
of \(\tau\FT=\SI{170}{\femto\second}\) to positively chirped pulses of
\(\SI{500}{\femto\second}\) (corresponding to 50--145 cycles) in 106
steps. The acquisition time for one image was around
\SI{80}{\milli\second}, averaging around \SI{1600}{shots}.

The calibration of the pulse duration as a function of the
grating position was based on the peak intensity of the pulse. The
observed cut-offs of HHs~25--37 were mapped to a specific driving
intensity using the cut-off law
\begin{equation}\label{eq:Cut-Off}
  q\hbar\omega_{0}=3.17 \frac{e^2 I_0(\tau)}{2 c \epsilon_0
    m\omega_{0}^2} + \Ip = 3.17\Up + \Ip,
\end{equation}
where $q$ is the harmonic order, $\hbar$ is the reduced Planck
constant, $\omega_0$ is the angular central frequency of the driving
laser, $\Ip$ is the ionization potential of argon, $e$ and $m$ are the
charge and the mass of electron, $\epsilon_0$ is the permittivity of
vacuum and $I_0(\tau)$ is the peak laser intensity for the pulse of
duration $\tau$, at the centre of the driving field. The laser peak
intensity is taken to vary as
\begin{equation}\label{eq:Itau}
  I_0(\tau)=I_0(\tau\FT) \frac{\tau\FT}{\tau},
\end{equation}
where the peak intensity at FT-limited duration \(I_0(\tau\FT)\) is on
the order of \intensity{e14}.  The laser beam with a diameter of
\SI{3.5}{\milli\meter} was focused by an achromatic lens with a focal
length \(\SI{100}{\milli\meter}\), resulting in a beam waist of
\(\SI{18}{\micro\meter}\) (estimated using Gaussian optics). As
generating medium argon gas was used, supplied through an open-ended,
movable gas nozzle with \(\SI{90}{\micro\meter}\) inner diameter. The
relative position of the nozzle and the laser focus was such that
phase matching allowed the observation of both short and long
trajectory harmonics~\cite{Salieres1995,Salieres2001S}.

The generated harmonic emission was analysed by a flat-field
grazing-incidence XUV spectrometer (based on Hitachi Grating 001-0639,
with the nominal line-spacing of \SI{600}{lines\per\milli\meter}). The
grating diffracted and focused the harmonics in the dispersive plane
and reflected them in the perpendicular direction onto a
\(\SI{78}{\milli\meter}\) diameter microchannel plate (MCP, Photonis),
which was imaged by a CCD camera (Allied Vision Technologies, Pike
F-505B with a pixel size of \SI{3.45 x 3.45}{\micro\meter}; the
resolution was set to \SI{2000 x 2000}{pixels} and the dynamic range
to \SI{14}{bits}). This arrangement allowed to study the spectral
content of the emission as well as the divergence of the individual
harmonics. The HH spectra were recorded for 106 positions of the pulse
compressor grating, see figure~\ref{fig:Scheme}. A more detailed
description of the setup can be found in \cite{Lorek2014}.


\section{Experimental results}
\label{sec:exp-results}
\begin{figure}
  \begin{center}
    \includegraphics[scale=0.65]{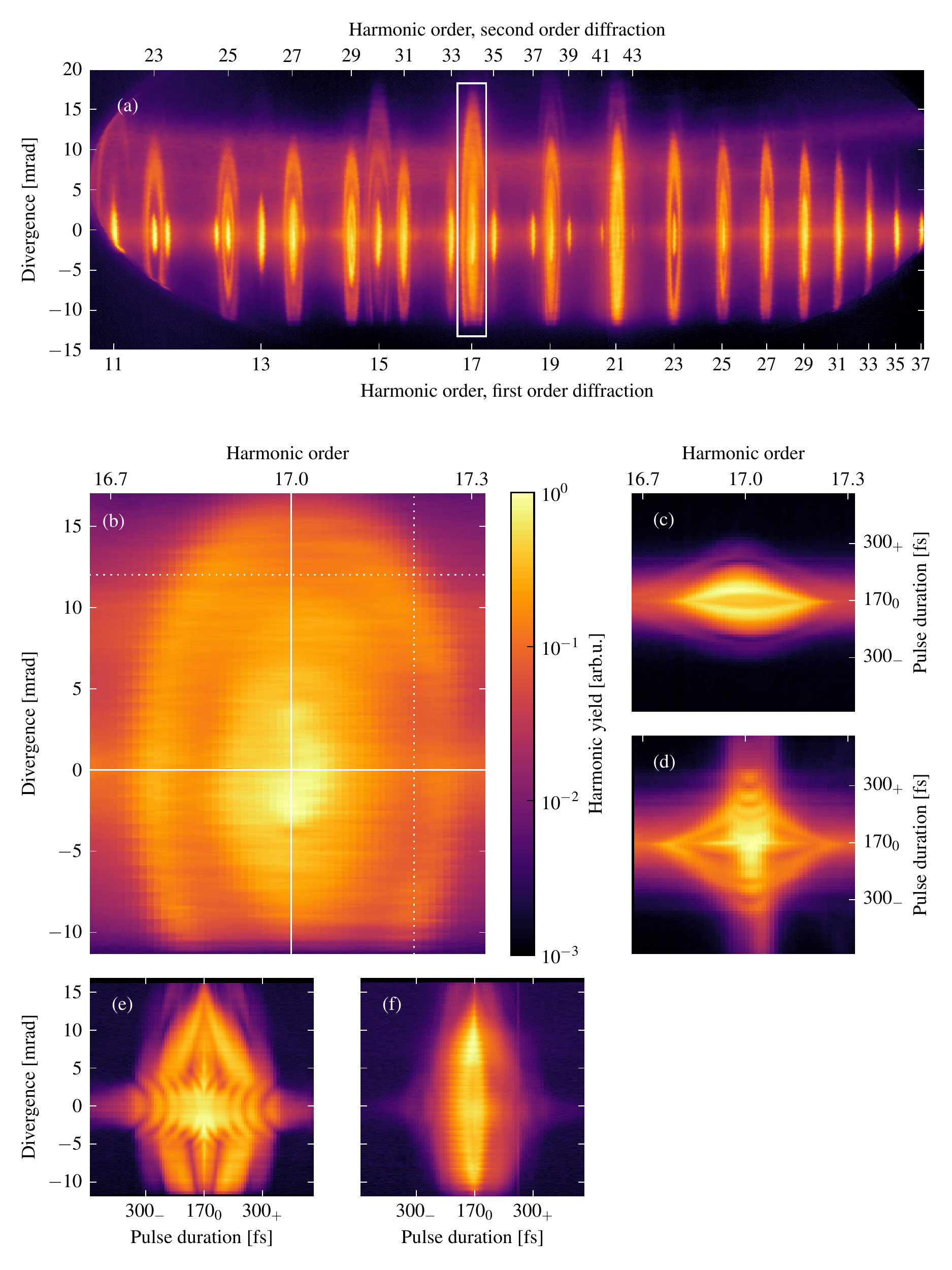}
  \end{center}
  \caption{\label{fig:IntroductionLineouts} { (a) Observed HHs as
      recorded by MCP under driving with transform-limited pulses,
      numbers indicate the spectral position of the first and second
      order diffraction of HHs for \(q=11-43\), the rectangle shows
      the area, for which an example of the analysis is given. (b)
      Magnified image of the area around HH17. The solid lines
      indicate where the lineouts on-axis and on the central harmonic
      energy, respectively, were made. Similarly, the dotted lines
      indicate where the lineouts off-axis and off the central
      harmonic energy were made. (c) shows the off-axis spectral
      lineouts corresponding mainly to the long trajectory
      contribution, while (d) shows the on-axis spectral lineout with
      a clear QPI pattern. Similarly, (e) shows the on-centre spatial
      lineout with a clear QPI pattern and (f) is a spatial lineout
      covering mainly the long trajectory contribution.
      \(300_{+/-}\)~fs means \SI{300}{\femto\second} pulse duration
      with positive/negative chirp; \(170_0\)~fs means FT-limited
      duration. The colour scale is logarithmic and is the same in all
      figures throughout the article, unless stated otherwise.}}
\end{figure}

A typical image of HHs on an MCP is displayed in
figure~\ref{fig:IntroductionLineouts}(a) for the case of a FT-limited
driving pulse. The HHs are both spatially and spectrally divergent,
with clear ring structures appearing around a strong, narrow central
structure. The off-axis structures are attributed to the long QP only,
whereas the on-axis structures contain both QPs. However, the on-axis
structures do not show to any visible modulation in a single
spectrum. To reveal the on-axis interference, the acquired HH
spatial--spectral profiles for 106 different values of the chirp
parameter were analysed by plotting different lineouts of the images
as a function of the driving pulse duration.  The spatial--spectral
profile of \(q=17\) is shown magnified in
figure~\ref{fig:IntroductionLineouts}(b). The horizontal axis (and
lineouts) correspond to the spectral variation, whereas the vertical
axis (and lineouts) to the spatial variation. The lines represent
regions of interest, from which subfigures (c)--(f) are extracted. (d)
and (e) are the spectral and spatial lineouts of the central area of
the generated harmonics with contribution from both the long and short
trajectories, while the off-axis lineouts (c) and (f) show mainly
behaviour of the long trajectory contribution, therefore lacking
interference between the two trajectories (however, long--long
interference remains). In the following analysis we focus on the
on-axis areas, where short--long QPI patterns are apparent, as in (d)
and (e). The spatial and spectral lineouts for other orders are
presented in sections \ref{sec:spatial-model} and
\ref{sec:spectral-model}. We return to the off-axis interference
structures in section~\ref{sec:long--long}.

\section{Mathematical model}
\label{sec:math-model}
To explain and analyse the observed QPI, we have developed a
mathematical model based on the interplay of the HH contributions from
different QPs; similar to the simple model of~\cite{Gaarde1999}.  We
concentrate on the two first trajectories, the so called short ($s$)
and long ($l$).  The main difference is that, since the long
trajectory spends more time in the continuum, it acquires more phase
which leads to larger divergence, both spatially and spectrally. This
phase is labelled $\Phi_{s/l}$. Using the SFA, the dipole phase can be
calculated by integrating the semi-classical action
\cite{Lewenstein1994,Varju2005JoMO} (in atomic units):
\begin{equation}
\label{eq:semi-classical-action}
\Phi_{s/l}(t_i,t_r,\vec{p})=q\omega_0 t_r -
\int_{t_i}^{t_r}\mathrm{d}t\left\{
  \frac{[\vec{p}+\vec{A}(t)]^2}{2}+\Ip
\right\},
\end{equation}
where the trajectory of the electron is defined by its ionization
time, \(t_i\), return time, \(t_r\), and momentum
\(\vec{p}\). \(\vec{A}(t)\) is the vector potential of the driving
field.
\begin{figure}[b]
  \centering
  \includegraphics{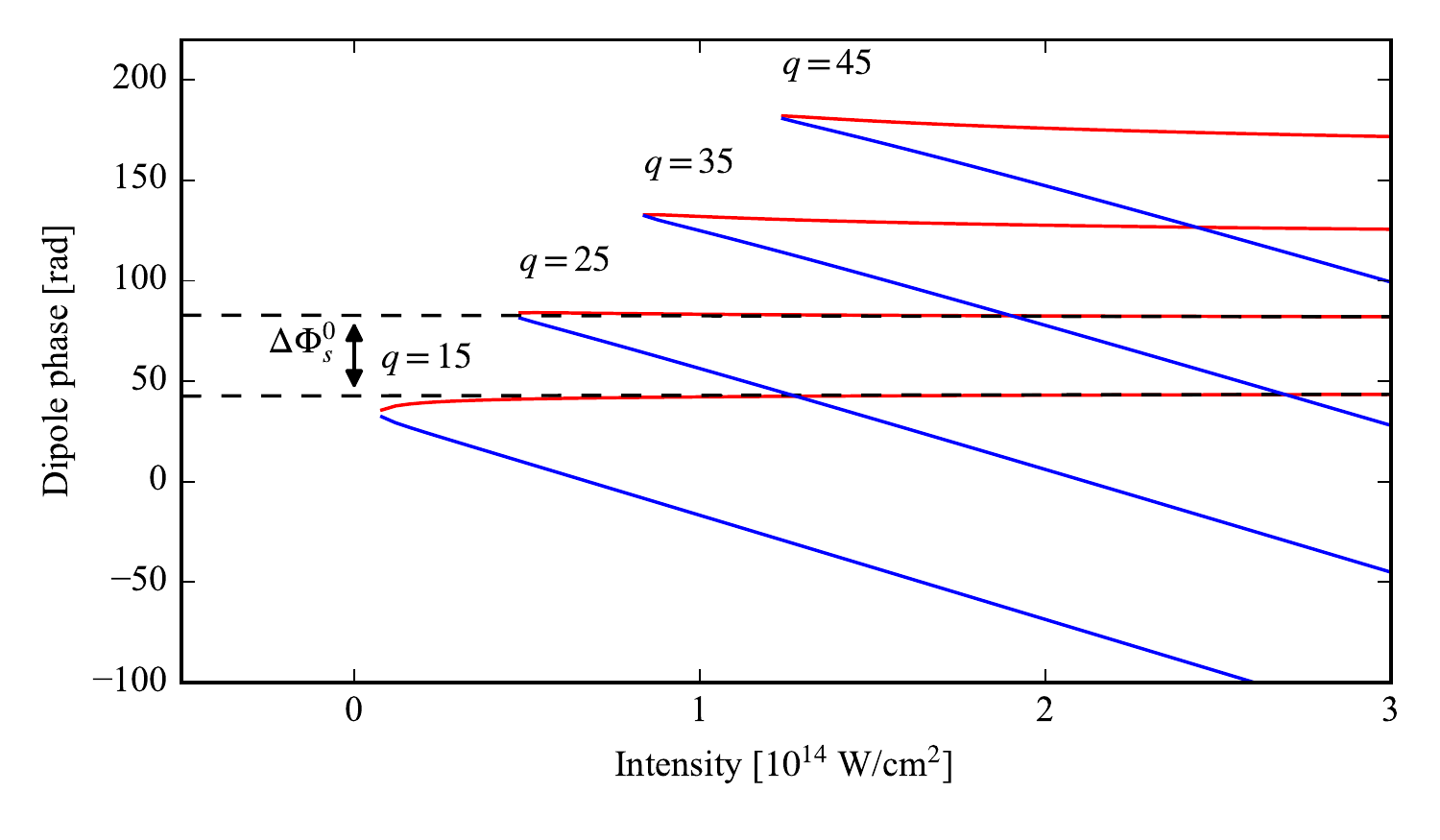}
  \caption{Dipole phase as a function of intensity, for different
    high-order harmonics of \(\lambda=\SI{1030}{\nano\meter}\),
    calculated using \eqref{eq:semi-classical-action}. The red (blue)
    lines correspond to the short (long) trajectories. For the
    intensities used in the experiment
    (\SIrange{4e13}{1e14}{\watt\per\centi\meter\squared}), the curves
    are well approximated by \eqref{eq:DipolePhase}. The two dashed
    lines are fits to the asymptotes of the red lines, i.e.\  they are
    not perfectly horizontal. The slopes of these fits are
    \(-\alpha_s\).}
  \label{fig:dipole-phase-offset}
\end{figure}
In figure~\ref{fig:dipole-phase-offset},
\eqref{eq:semi-classical-action} is plotted for a few different
harmonic orders, using the experimental conditions of the present
work. The ionization time \(t_i\) and the return time \(t_r\) are
found by solving Newton's equations of an electron in an electric
field. The general behaviour of the phases as a function of intensity
leads us to the following approximate expression:
\begin{equation}\label{eq:DipolePhase}
  \Phi_{s/l}(r,z,t) = \Phi^0_{s/l} + \alpha_{s/l} I(r,z,t),
\end{equation}
where \(\Phi_{s/l}^0\) is a phase offset and $\alpha_{s/l}$ are slopes
of the phases as function of the intensity \cite{Gaarde1999}. This
adiabatic model is valid for the experimental conditions of the
present work~\cite{Murakami2005a}.

In our simple model, we assume a tight-focus geometry with a small
interaction volume and we only consider HH generated in the focal
plane $z=0$.

\subsection{Divergence}
\label{sec:spatial-model}
To model the behaviour of the harmonics along the divergence axis as
the pulse duration varies, the emission from the two trajectories is
approximated as a sum of Gaussian beams. Such beams can be propagated
to the far field analytically (in the paraxial approximation), and the
geometrical properties necessary are determined from the experimental
conditions. In~\ref{app:spatial-model}, the full derivation of the
divergence model can be found. The main result is that the total far
field can be written in cylindrical coordinates as
\begin{equation}
  E_{\textrm{detector}}(r,z)=E_{s}(r,z)+E_{l}(r,z),
\end{equation}
where
\begin{equation}
  E_{s/l}(r,z)=C_{s/l}I_0^{\frac{n}{2}}(\tau)W(z)\exp{\left[-\im G(r,z;r_0,z^R)+\im\Phi_{s/l}(r,z)\right]}.
\end{equation}
\(C_{s/l}\) are weights for the trajectories, \(n\) is a nonlinearity
parameter, \(W(z)\) and \(G(r,z;r_0,z^R)\) are functions depending on
the geometry as well as the ionization process, whereas \(\Phi_{s/l}\) is
the phase in \eqref{eq:DipolePhase}, which only depends on the atomic
properties. \(r_0\) is the beam waist (\SI{18}{\micro\meter}) and
\(z^R\) the effective Rayleigh range (\(\sim\)\SI{1}{\milli\meter}).

\begin{figure}[bth]
  \centering
  \includegraphics {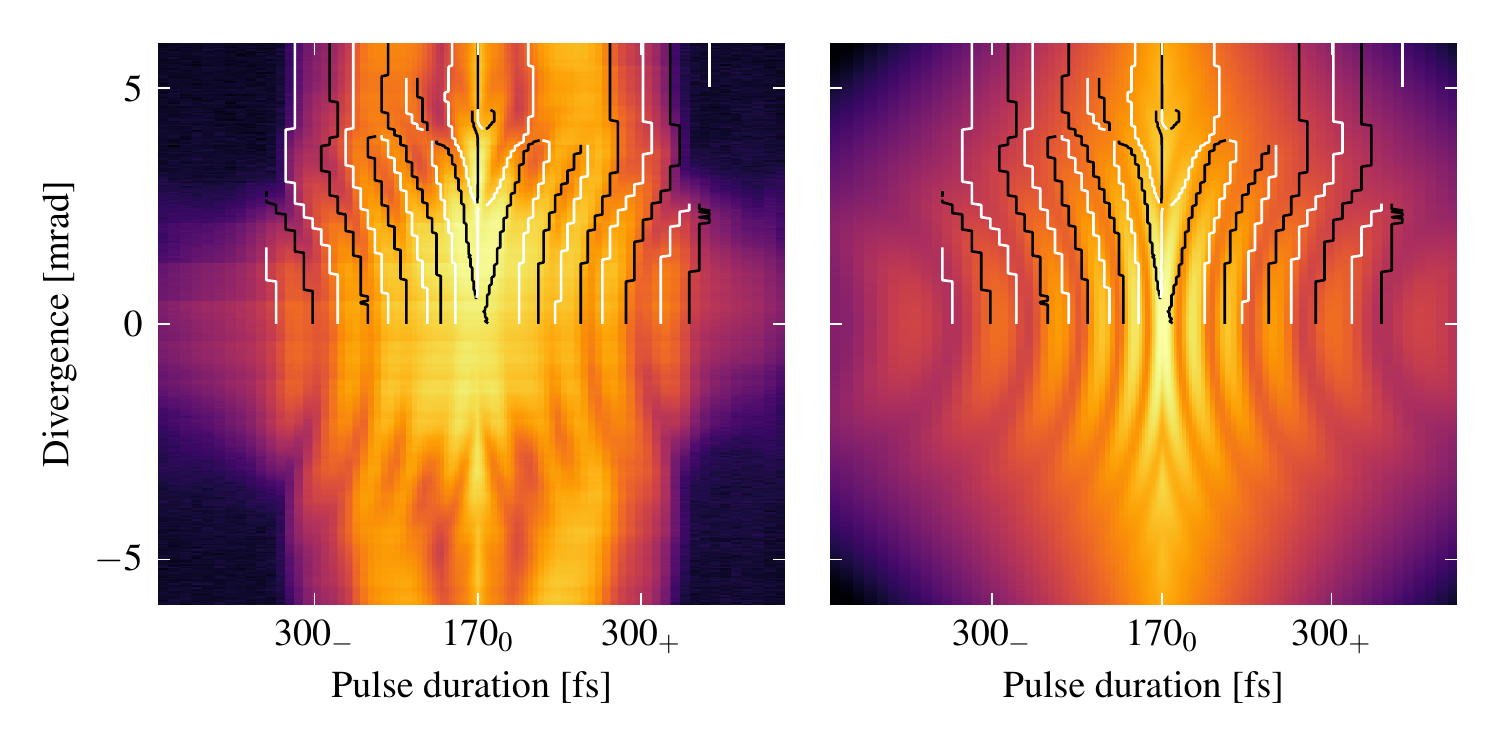}
  \caption{Illustration of the fitting procedure for retrieval of the
    values of \(\alpha_{s/l}\) for the different harmonics. In the left
    panel the experimental data are shown with the maxima (white
    lines) and minima (black lines) extracted. In the right panel, the
    coloured map is the result of the mathematical model described in
    section~\ref{sec:spatial-model}, while the lines are the same as
    in the left panel. \(300_{+/-}\)~fs means \SI{300}{\femto\second}
    pulse duration with positive/negative chirp; \(170_0\)~fs means
    FT-limited duration.}
  \label{fig:fitting-procedure}
\end{figure}

The spatial profiles of the generated HH beams are calculated for
\(q=11-37\), using the variation of the pulse duration $\tau$ as in
the experiment. The experimental input values are $\lambda$, $r_0$,
$z$ and $I_0(\tau)$ [determined using~\eqref{eq:Cut-Off}], whereas
unknown parameters are $\Phi^0_s$,$\Phi^0_l$,$\alpha_{s}$,
$\alpha_{l}$, $n$ and the ratio $C_{l}^2:C_{s}^2$. In our procedure we
neglect the influence of phase offset difference $|\Phi^0_s-\Phi^0_l|$
-- it influences only the phase of the fringe pattern (with 2$\pi$
periodicity), but not the shape. The procedure for retrieving the
values of \(\alpha_{s/l}\) for the different harmonics is the
following: 1) The experimental spatial lineouts are normalized
separately for each harmonic, 2) positions of interference maxima and
minima are determined [shown as the white and black lines overlaid in
figure~\ref{fig:fitting-procedure}] in order to highlight the shape of
the interference pattern, 3) the parameters of the model are then
fitted such that the frequency of the fringes and their curvature in
the model match that of the experiment (see figure
\ref{fig:fitting-procedure} for \(q=17\)).
The phase difference \((\alpha_l-\alpha_s)I\) between the two
trajectories can by itself explain the observed frequency of the
fringes, on-axis. However, we have more information available in that
the fringe pattern has a curvature, which allows us to retrieve not
only the difference between \(\alpha_{l/s}\), but also their absolute
values. This is because the curvature of the fringes depends on mean
value \((\alpha_l+\alpha_s)/2\) as well as on the difference
\(\alpha_l-\alpha_s\), which is why an iterative fit has to be
made. The contrast and the overall intensity of the divergence
pattern, are mainly affected by nonlinearity parameter $n$ and by the
ratio of long and short trajectory contribution $C_{l}^2:C_{s}^2$.

For \(q=15-21\), automated fitting of the model to the experimental
data could be done, while for the higher harmonics, a visual fit was
the only option, since the interference signal was too weak for these
harmonics. The retrieved values of the parameters are listed in
table~\ref{tab:SimulatedSpatial}.

\begin{figure}[h]
  \centering
  \subfigure[Experiment.]{%
    \includegraphics[width=7.4cm]{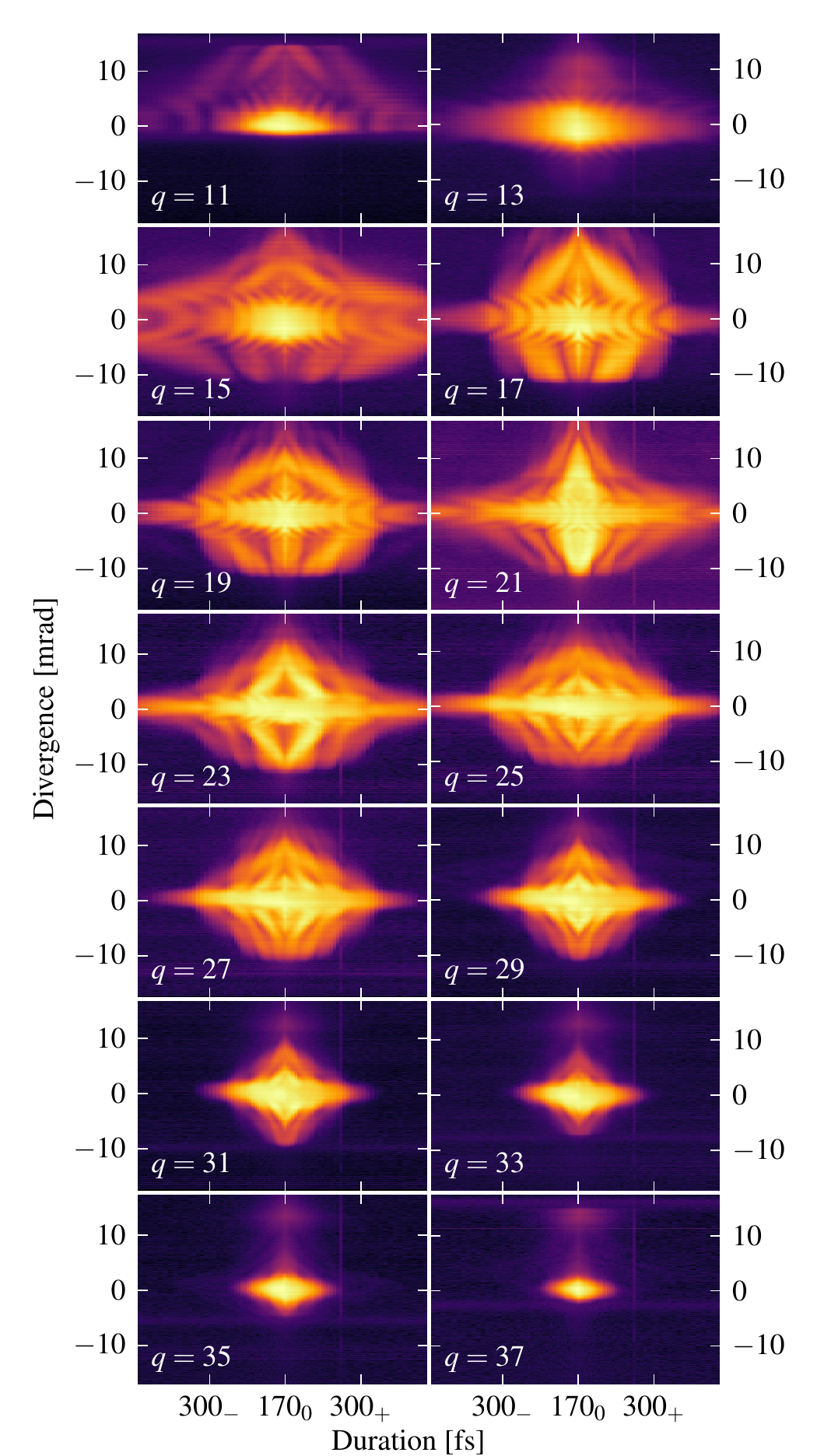}
    \label{fig:SpatialExp}}
  \quad
  \subfigure[Model.]{%
    \includegraphics[width=7.4cm]{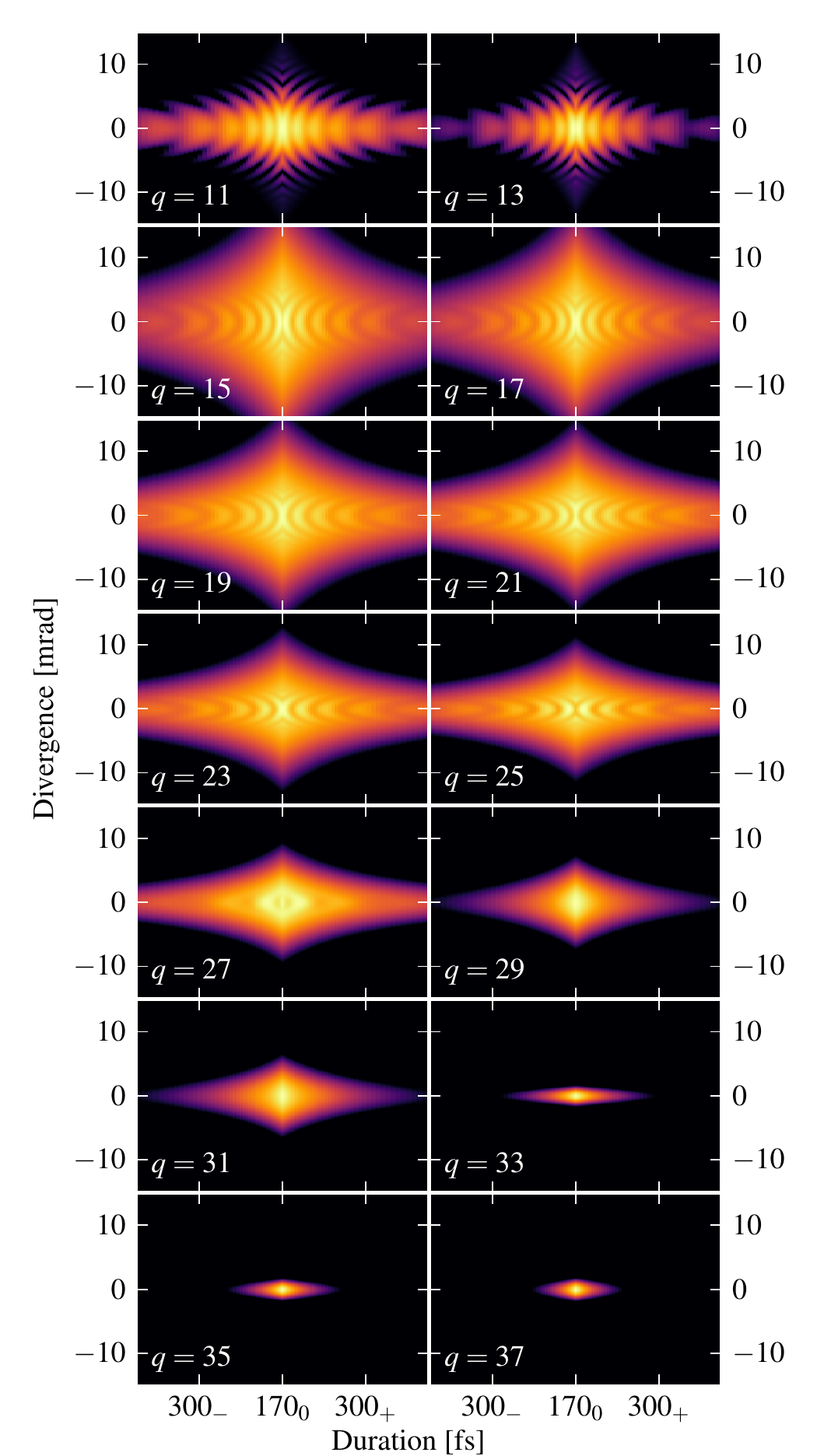}
    \label{fig:SpatialModel}}
  \caption{\label{fig:SpatialLineouts} Comparison of experimental and
    modelled data for variation of divergence profiles with gradually
    varied duration of the driving pulses for \(q=11-37\). In the left
    part of the images the driving pulses are negatively chirped,
    whereas in the right part are positively chirped. The signal for
    the low harmonic orders, especially apparent for \(q=11\), is
    limited to positive divergence by the shape of the MCP.}
\end{figure}

In figure~\ref{fig:SpatialModel}, the far-field divergence profiles as
simulated by the spatial model are shown for the values of
\(\alpha_{s/l}\) as extracted by the fitting procedure described above
(these values are presented in table~\ref{tab:SimulatedSpatial}). To
be noted is that the model manages to reproduce the hyperbolic fringes
visible for divergences smaller than \(\sim \SI{5}{\milli\radian}\);
however, the prominent ring structure visible for larger divergences
are not reproduced by this model. As will be discussed in more detail
in section~\ref{sec:long--long}, the rings arise when including
higher-order terms than parabolic in the phase curvature. The ring
structure in a given harmonic depend on higher-order corrections to
the harmonic phase beyond the simple linear dependence on the
intensity with phase coefficient $\alpha$. In particular, the far-field
radiation pattern for the long-trajectory harmonics consists of
interfering contributions from parts of the near-field where these
harmonics belong to the plateau and parts of the near-field where
these harmonics belong to the cut-off. Such contributions to the same
harmonic will have different values of $\alpha$.

%
%
\begin{table}[h]
  \caption{\label{tab:SimulatedSpatial}Values of parameters used in
    the divergence model [see figure~\ref{fig:SpatialModel}] for the
    simulation of the spatial profiles of generated HH.}
  \begin{indented}
  \item[]\scriptsize
    \begin{tabular}{@{}l*{15}{>{\!\!$}r<{$}}}
      \(q\)                                                                       & 11 & 13& 15&17&19 & 21& 23& 25&27&29 & 31 & 33& 35&37\\
      \br
      \(\alpha_{s}/(\SI{e-14}{\centi\meter\squared\per\watt})\)               & -10 &-11&-10&-10&-10&-9&-7&-5&5&15&15&18&20&24\\
      \(\alpha_{l}/(\SI{e-14}{\centi\meter\squared\per\watt})\)         &50&50&51&50&43&43&40&38&35&35&33&31&28&24\\
      $n$                                                                     &6&  8&   7&  7&    6& 6& 6& 6& 7&10&10& 11&14&17\\
      $C_{l}^2:C_{s}^2$                                                   &2& 2& 100&100&100&100&100&100&100&100&100&30&50&5 \\

    \end{tabular}
  \end{indented}
\end{table}

\subsection{Spectrum}
\label{sec:spectral-model}
To understand the variation of the spectral profile as a function of
the driving field chirp, it is important to model the temporal
behaviour of the harmonic generation, and particularly its response to
change in instantaneous frequency. The detailed derivation can be
found in~\ref{app:spectral-model}. The main result this time is that
the field contribution for a given harmonic from the short/long
trajectory can be written
\begin{equation}\label{eq:EfieldSpectral}
  E_{s/l}(t) = C_{s/l}
  I^{\frac{n}{2}}(t)\exp\left[\im q\omega_0+
    \im \frac{qb(\tau)}{2}t^2
    +\im\alpha_{s/l} I(t)+\im\Phi^0_{s/l}\right].
\end{equation}

The far-field spectra of the generated HHs are computed as the Fourier transform
of the sum of the fields generated by the short and long trajectory
contributions
\begin{equation}
  S(\omega)=\mathcal{F}[E_s(t)+E_l(t)],
\end{equation}
and the intensity in the far field is given by
\begin{equation}
  I(\omega) \propto |S(\omega)|^2.
\end{equation}

\begin{figure}[h]
  \centering
  \subfigure[Experiment.]{%
    \includegraphics[width=7.4cm]{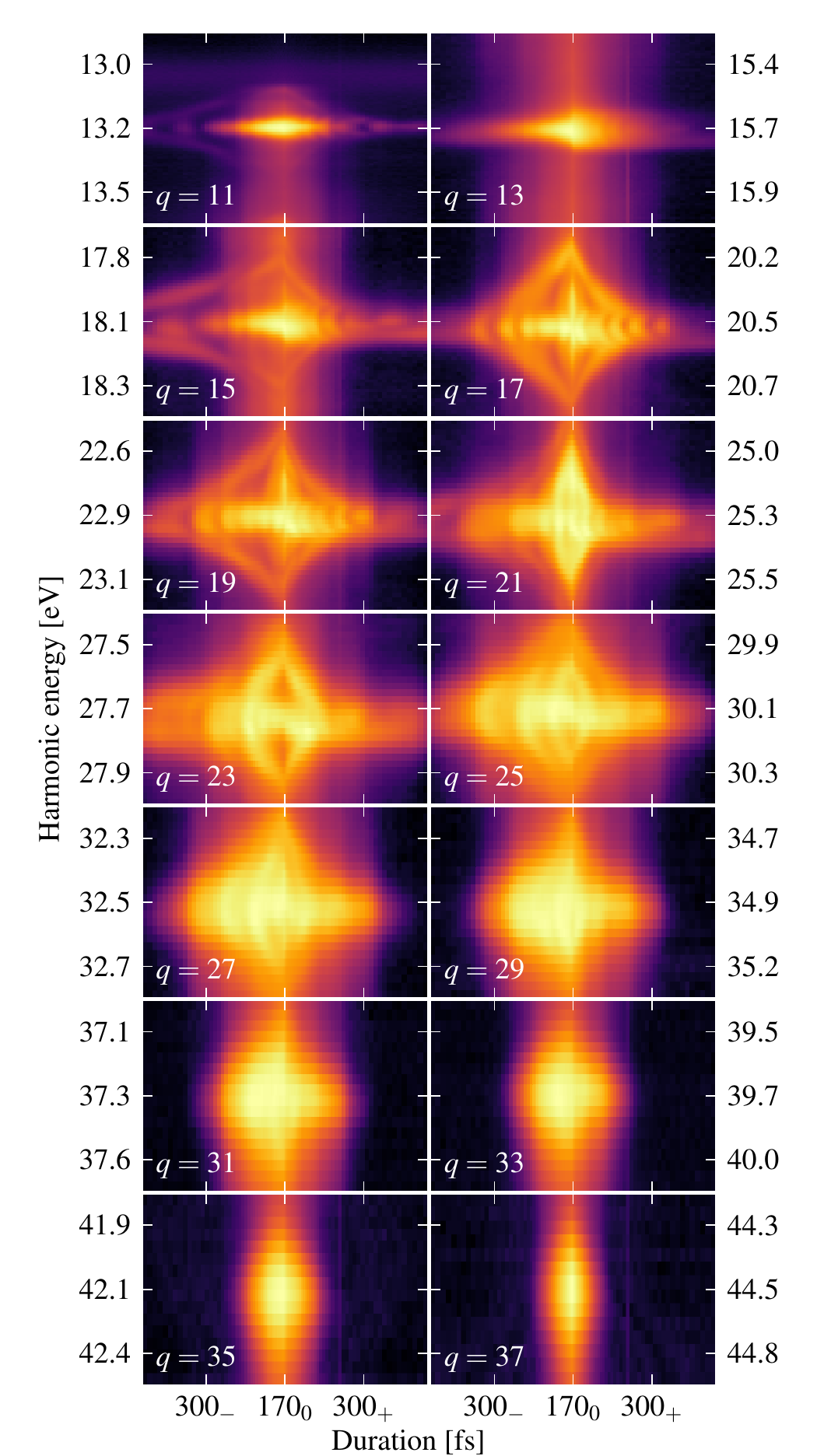}
    \label{fig:SpectralExp}}
  \quad
  \subfigure[Model.]{%
    \includegraphics[width=7.4cm]{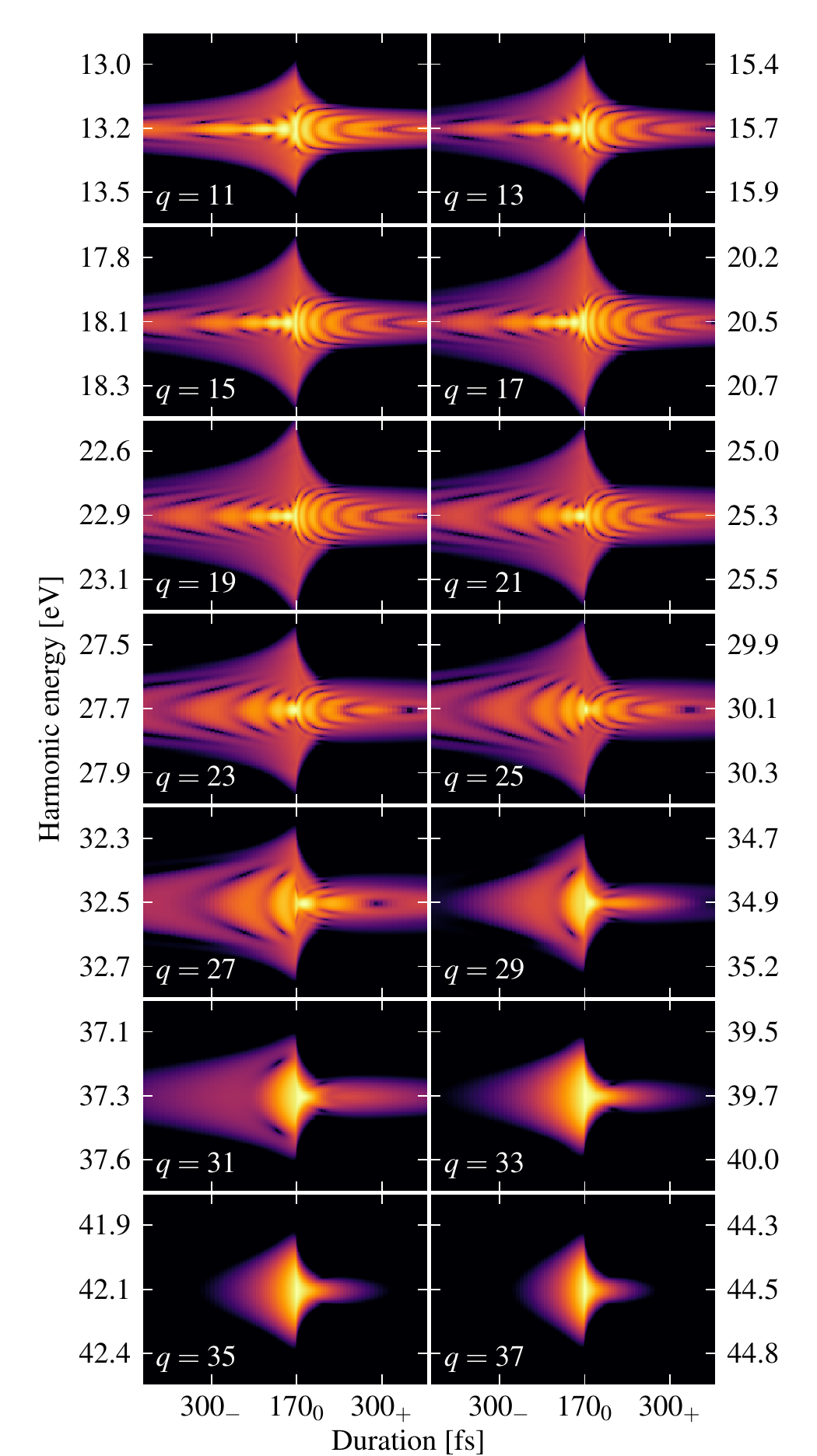}
    \label{fig:SpectralModel}}
  \caption{\label{fig:SpectralLineouts} Comparison of experimental and
    modelled data for variation of HH spectra with gradually varied
    duration of the driving pulses for \(q=11-37\). In the left part
    of the images the driving pulses are negatively chirped, whereas
    in the right part they are positively chirped. As is explained
    section~\ref{sec:spectral-model}, the fit of the spectral model to
    the data was purely visual, matching the the amount of fringes and
    their positions.}
\end{figure}

The simulated spectra are shown in
figure~\ref{fig:SpectralLineouts}(b). The experimental input
parameters are the same as for the spatial profile simulation and the
retrieved values of the parameters are listed in
table~\ref{tab:SimulatedSpectra}. Again, \(\Phi_{s/l}^0\) is assumed
to be zero, since it only influences the phase of the interference
fringes, but not their shape. When determining the values of
$\alpha_{s/l}$, the main attention is given to the width of the
measured spectra and to the curvature of the QPI fringes. The values
were found by a pure visual fit of the model to the experimental data,
no automated fitting was employed. The $n$ was kept same as in the
simulation of the divergence lineouts, while the ratio
$C_{l}^2:C_{s}^2$ had to be decreased, due to the fact that only the
middle part of the divergence cone is evaluated [see
figure~\ref{fig:IntroductionLineouts}(d)]. In this cone, the relative
contribution of the short trajectory is much stronger than when a
broad divergence region is considered, leading to a lower ratio
$C_{l}^2:C_{s}^2$.

The asymmetric behaviour of the central part of the spectra with
respect to the chirp parameter (clearly apparent in
figure~\ref{fig:SpectralExp} for \(q=13,15,17\)), enables us to
determine negative values of \(\alpha_s\) in the region below threshold and
in the plateau. It is possible to make this identification, since the
central part of the spectrum is dominated by the short trajectory
contribution. The instantaneous frequency of the generated HH field is
described by \eqref{eq:omegaHH} -- the time derivative of the argument
of~\eqref{eq:EfieldSpectral}. For negatively chirped pulses, the chirp
of the driving pulse [the second term in \eqref{eq:omegaHH}] has the
same sign as the chirp introduced by the dipole phase [third term in
\eqref{eq:omegaHH}] and a broad spectrum of frequencies is
generated. In contrast, when the pulse is positively chirped, the
second and third terms have opposite signs and partly compensate each
other, with a narrower spectrum as the result. The negative sign of
\(\alpha_s\) for the low orders leads to this compensation occurring for
negatively chirped pulses (the left side of the spectra in
figure~\ref{fig:SpectralLineouts}), while for higher orders, the
compensation occurs for positively chirped pulses (the right side of
the spectra in figure~\ref{fig:SpectralLineouts}). The sign change
occurs around harmonic 23, where the narrowest spectrum of short
trajectory is found for FT-limited pulses. For the long trajectory,
all \(\alpha_l\) are positive, such that the compensation always occurs for
positively chirped pulses.

%
\begin{table}[b]
  \caption{\label{tab:SimulatedSpectra}Values of parameters used in
    the spectral model [see figure~\ref{fig:SpectralModel}] for the
    simulation of the spectral profiles of generated HH.}
  \begin{indented}
  \item[]\scriptsize
    \begin{tabular}{l*{15}{>{\!\!$}r<{$}}}
      \(q\)                                                                       & 11 & 13& 15&17&19 & 21& 23& 25&27&29 & 31 & 33& 35&37\\
      \br
      \(\alpha_{s}/(\SI{e-14}{\centi\meter\squared\per\watt})\)               & -10 &-11&-9&-10&-8&-4&-2&4&10&14&16&20&20&20\\
      \(\alpha_{l}/(\SI{e-14}{\centi\meter\squared\per\watt})\)         &35&40&48&55&56&50&45&50&40&35&30&27&27&27\\
      $n$                                                                     &6&  8&   7&  7&    6& 6& 6& 6& 7&10&10& 11&14&17\\
      $2C_{l}^2:C_{s}^2$                                                   &1& 1& 1&1&1&1&1&1&1&1&1&1&1&1 \\

    \end{tabular}
  \end{indented}
\end{table}

\subsection{Dipole phase parameters}
\begin{figure}[h!]
  \centering
  \includegraphics{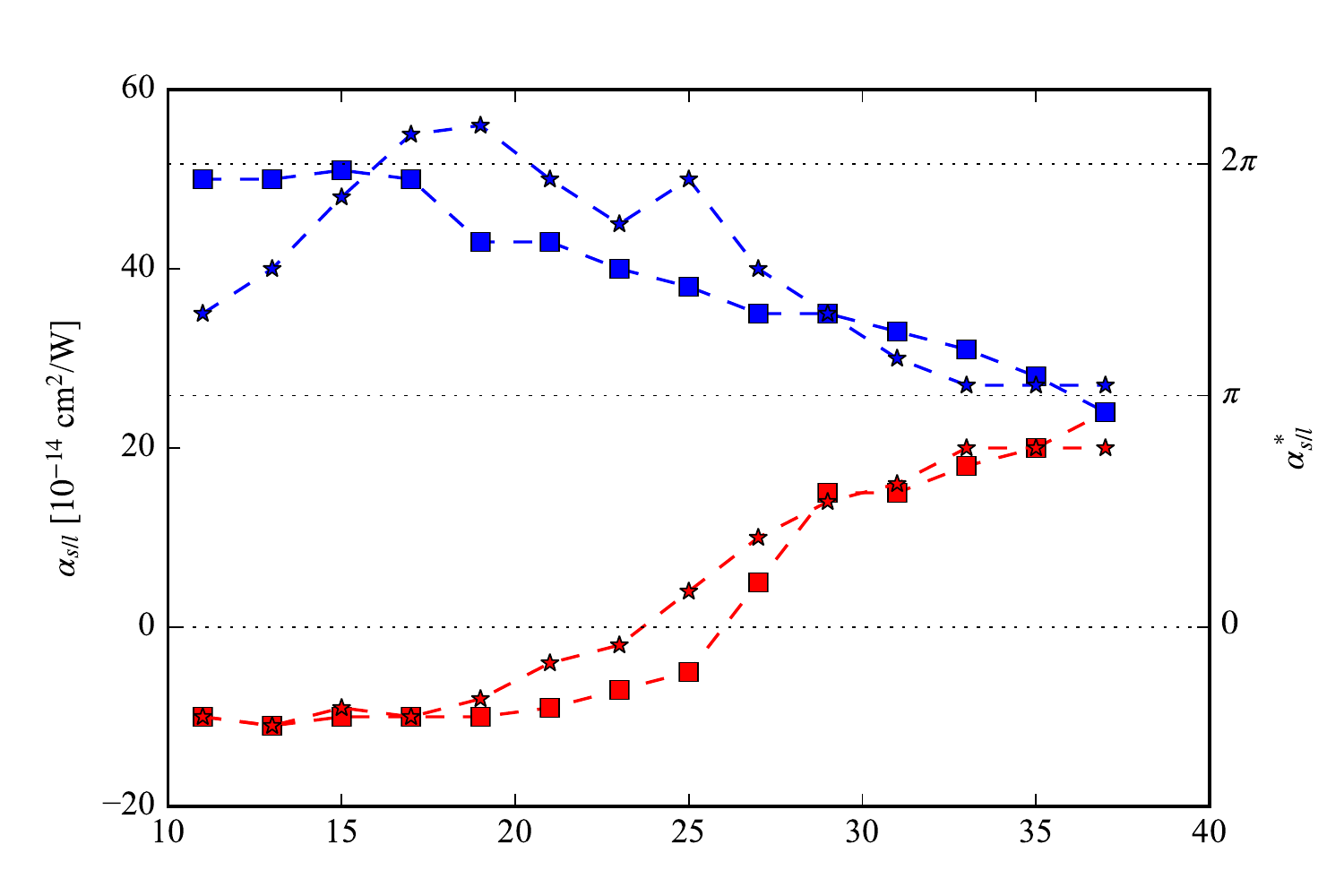}
  \caption{\label{fig:Alphas} Overview of the retrieved dipole phase
    parameters $\alpha_s$ and $\alpha_l$ from the divergence model (squares) and
    from the spectral model (stars), for long (blue) and short (red)
    trajectory, along with their estimated uncertainties. The
    right-hand scale is calculated according to
    equation~\eqref{eq:AlphaRed}.}
\end{figure}

In figure~\ref{fig:Alphas}, the retrieved values of \(\alpha_{s/l}\)
from both the divergence model and the spectral model, are shown. The
values of $\alpha_{s/l}$ predicted by different theoretical
calculations and retrieved for various experimental conditions
(driving wavelengths $\lambda$ and intensities $I$) can be compared,
by expressing them in dimensionless values $\alpha_{s/l}^*$ related to
the optical cycle of the driving pulse~\cite{Yost2009}:
\begin{equation}\label{eq:AlphaRed}
  \alpha_{s/l}^*=\frac{2c \epsilon_0 m \omega_0^3 \hbar}{e^2}\alpha_{s/l}.
\end{equation}
The theory predicts the values of \(\alpha_l^*\approx2\pi\) and
\(\alpha_s^*\approx0\) in the plateau region, with both values converging to
$\pi$ towards the cut-off \cite{Lewenstein1995,Varju2005JoMO}.
The errors in the parameters \(\alpha_{s/l}\) are difficult to
quantify. It is possible to make an estimate by comparing the values
extracted using the divergence model and spectral model. They should
in principle yield the same values, since they are both based on the
assumption that the phase can be written as stated in
\eqref{eq:DipolePhase}. However, since the spectral data are extracted
from a smaller part of the divergence cone than the spatial data, the
former are more sensitive to errors which could explain the larger
variation in the data. An estimate of the error is given by the mean
discrepancy between the two models, which is
\(\sim2.5\times\SI{e-14}{\centi\meter\squared\per\watt}\) for the
short trajectory and
\(\sim6\times\SI{e-14}{\centi\meter\squared\per\watt}\) for the long
trajectory.


\section{Quantum mechanical calculations}
\label{sec:tdse}
\begin{figure}[h!]
  \centering
  \includegraphics{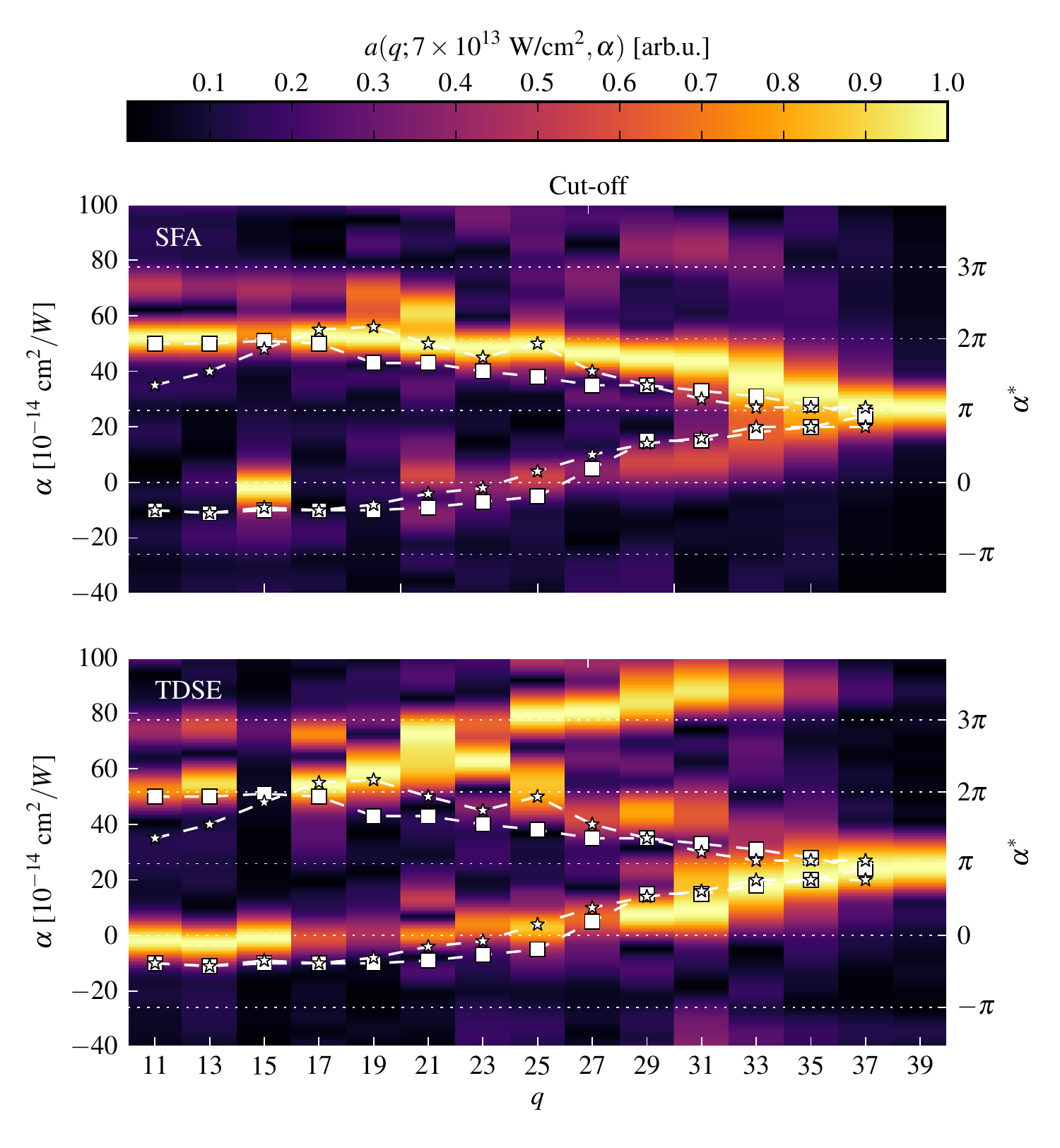}
  \caption[Quantum path distributions (QPDs) for the harmonics
  \(q=\)11--39]{Quantum path distributions (QPDs), normalized, for the harmonics
    \(q=\)11--39, calculated at the driving field intensity
    \(\intensity{7e13}\). Bright regions correspond to more likely
    values of \(\alpha\) for a certain harmonic order. The white lines
    correspond to the experimentally retrieved values of \(\alpha\),
    with the lower values belonging the short trajectories and the
    higher values to the long trajectory. The right-hand \(\alpha^*\)
    (the variable conjugate to \(I\)) scale is given in radians in
    accordance with \eqref{eq:AlphaRed}.

    In the SFA, the long trajectory is significantly more prevalent
    compared to the short trajectory, and this has been observed
    before~\cite{Balcou1999}. In contrast, the TDSE yields short and
    long trajectories of comparable weight, and even longer
    trajectories are visible; also this is a previously known
    result~\cite{Murakami2005}. The third trajectory has not been
    observed in the experiments, which might be due to the
    unfavourable phase matching conditions.}
  \label{fig:quant-alphas}
\end{figure}

For comparison, calculations of the HH yield are performed by
integrating the TDSE for a range of intensities using a newly
developed graphics processing unit (GPU) implementation of the
algorithm outlined in \cite{Schafer2009}. For a large range of
intensities, the time-dependent dipole acceleration \(a(t,I)\) of the
atom is computed, and the quantum path distributions (QPDs) are
extracted in the same manner as described in great detail in
\cite{Balcou1999}; first a Fourier transform is performed to get the
spectrum \(a(\omega,I)\) and subsequently, for each harmonic order
\(q\), a Gabor transform is performed along the intensity axis to
obtain the QPDs \(a(q;I,\alpha)\). In figure \ref{fig:quant-alphas}, the
QPDs leading to emission of the different harmonics are plotted in an
intensity range around the experimental intensity, along with the
experimentally retrieved values of \(\alpha_{s/l}\) as presented in tables
\ref{tab:SimulatedSpatial},~\ref{tab:SimulatedSpectra}. For
comparison, the same procedure is performed for the SFA; the main
difference is that the long trajectory is much more pronounced in the
SFA, whilst the TDSE also shows longer trajectories. In general,
though, they both agree well with the experimental results.

One important difference compared to the models presented above, is
that these calculations are performed at slightly lower intensity,
\(\intensity{7e13}\) as compared to \(\intensity{1e14}\). These
calculations are performed using a trapezoidal pulse shape, with
exactly this intensity, while in the experiment and the models, the
pulse shape is Gaussian, which naturally spans a distribution of
intensities, up to the nominal intensity, \(I_0(\tau\FT)\).


\section{Analysis of off-axis ring-like structures}
\label{sec:long--long}

\begin{figure}[h]
  \centering
  \includegraphics{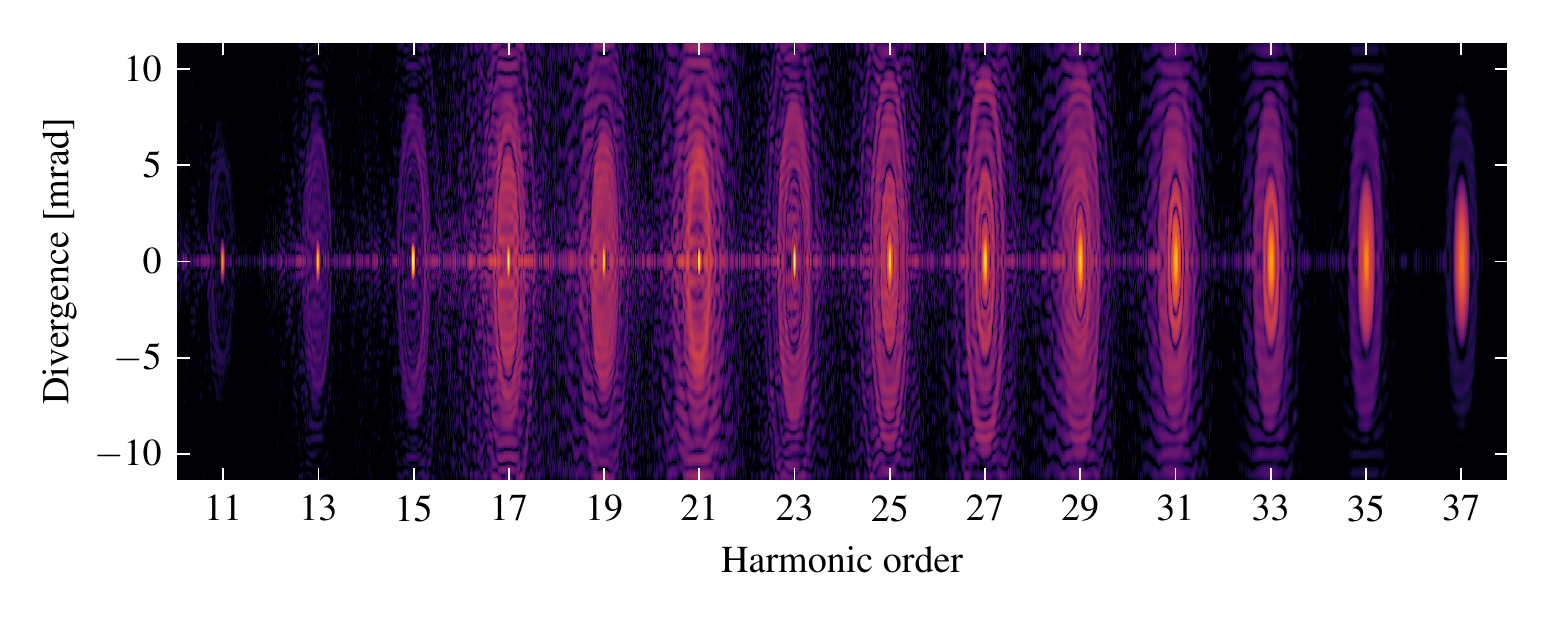}
  \caption{Theoretical far-field spectrum
    [cf. figure~\ref{fig:IntroductionLineouts}(a)], for the case for
    FT-limited driving pulse (\SI{170}{\femto\second}). The
    single-atom response of a set of atoms is calculated using the
    TDSE and propagated via a Hankel transform to the far-field.}
  \label{fig:theory-far-field-spectrum}
\end{figure}

In figure~\ref{fig:theory-far-field-spectrum}, a theoretical far-field
spectrum calculated for the parameters of the experiment is shown.
The time-dependent dipole acceleration \(a(t)\) is calculated by the
TDSE for a set of atoms in the focal plane. The collective emission is
propagated to the far-field, as is described below in
section~\ref{sec:far-field-propagation}. Qualitatively, the agreement
with the experimental spectrum in
figure~\ref{fig:IntroductionLineouts}(a) is very good; the appearance
of further spatial modulation can be attributed to the lack of
intensity averaging as is present in the experiment. Notably, the
ring-like structures off-axis (i.e.\ for divergences
\(\geq\SI{5}{\milli\radian}\)) are present, whereas they are missing in
the results of the Gaussian model in its parabolic phase approximation
as presented in figure~\ref{fig:SpatialModel}. This can be understood
as follows: The harmonic emission can be written as
\begin{equation}
  \label{eq:hhg-model}
  E(r,t) = A(r,t)\exp[\im\Phi(r,t)],
\end{equation}
where \(A(r,t)\) is the amplitude and \(\Phi(r,t)\) the phase, both
dependent on the location and time of emission. If we assume we can
divide the emission into different contributions from different
harmonic orders \(q\) and different trajectories \(j\), we get
\begin{equation}
  \label{eq:hhg-traj-model}
  E(r,t) = \sum_{qj}A_q(r,t)\exp[\im\Phi_{qj}(r,t)],
\end{equation}
with
\[\Phi_{qj} = \Phi^0_{qj} + \underbrace{\pd{\Phi_{qj}}{I}}_{\alpha_{qj}}I(r,t) +
  \Ordo\left(\pdd{\Phi_{qj}}{I}\right),\]
and
\[\Phi^0_{qj} = q\Phi_0(t) + \Phi_{qj}(I_0)\]
contains the phase of the fundamental \(\Phi_0(t)\) and the atomic dipole
response at the peak of the field. In the Gaussian model, the
amplitude \(A_q(r,t)\) is of the form \(I^{\frac{n}{2}}(r,t)\), where
the fundamental field intensity is given by
\begin{equation}
  \begin{aligned}
    I(r,t)
    &= I_0(\tau)\exp\left(-\frac{t^2}{2\tau^2}\right)\exp\left(-\frac{r^2}{2r_0^2}\right)\\
    &= I_0(\tau)\exp\left(-\frac{t^2}{2\tau^2}\right)\left[1-\frac{r^2}{2r_0^2} + \frac{r^4}{8r_0^4} + \Ordo(r^6)\right].
  \end{aligned}
\end{equation}
The normal approximation is to neglect terms of
\(\Ordo\left(\pdd{\Phi_{qj}}{I}\right)\) and higher. Furthermore, it is
only possible to analytically propagate the emission to the far-field
if the radial profile of the intensity in the phase is approximated up
to second order in \(r\). By including higher-order terms of the
spatial profile through a numerical far-field transform, ring
structures appear in the far-field amplitude (see
figure~\ref{fig:far-field-propagation}). It is not enough, however, to
fully explain the off-axis behaviour of the interference rings -- the
long trajectory also probes a wider range of intensities, also those
for which a certain harmonic would be considered to be in the cut-off
regime. This means we cannot ignore the influence of
\(\pdd{\Phi_{qj}}{I}\) and higher-order terms in the expansion of the
phase with respect to the intensity. The effects of these
considerations will be briefly surveyed below.

\subsection{Adiabatic model}
The Gaussian model in its simplest form does not explain the correct
behaviour of the off-axis emission. To find the missing link, we
employ an adiabatic model, where instead of assuming the form
\eqref{eq:hhg-traj-model}, we opt for something in-between
\eqref{eq:hhg-model} and \eqref{eq:hhg-traj-model}:
\begin{equation}
  \label{eq:adiabatic-model}
  A(r) = \sum_q a[q;I(r)],
\end{equation}
that is, we still decompose the emission into different harmonic
orders, but it is not trajectory-resolved anymore.  \(a[q;I(r)])\) can
be the dipole acceleration moment for harmonic order \(q\) as
calculated using the TDSE, in the manner described in
section~\ref{sec:tdse}, or the dipole spectrum from the SFA. The model
is adiabatic inasmuch it does not consider the temporal intensity
variation of the driving pulse, but only the spatial intensity
variation \(I(r)\) at the peak of the pulse. Furthermore, only
emission from the focal plane is considered.

\subsection{Exact far-field propagation}
\label{sec:far-field-propagation}
The far-field amplitude of the emission is found by propagation. In
cylindrical coordinates and cylindrical symmetry, this is given by
\cite{Goodman1996}:
\begin{equation}
  U_0(\rho) = -\im\frac{k}{2\pi z}
  \exp(\im kz)\exp\left(\im\frac{k}{2z}\rho^2\right)
  \mathcal{H}_0\{A(r)\}(k\rho/2\pi z),
\end{equation}
where \(\mathcal{H}_0\{A(r)\}\) is the zeroth-order Hankel transform
of the near-field radial amplitude \(A(r)\), \(r\) is the near-field
radial coordinate, \(\rho\) is the far-field radial coordinate, \(k\) is
the wavevector \(q2\pi/\lambda\) (\(q\) is the harmonic order and
\(\lambda\) is the fundamental wavelength) and \(z\) is the propagation
distance. The Hankel transform is computed numerically using the
algorithm presented in \cite{Guizar-Sicairos2004JotOSoAA}.

\subsection{Off-axis interference structures}
\begin{figure}[h]
  \centering
  \includegraphics{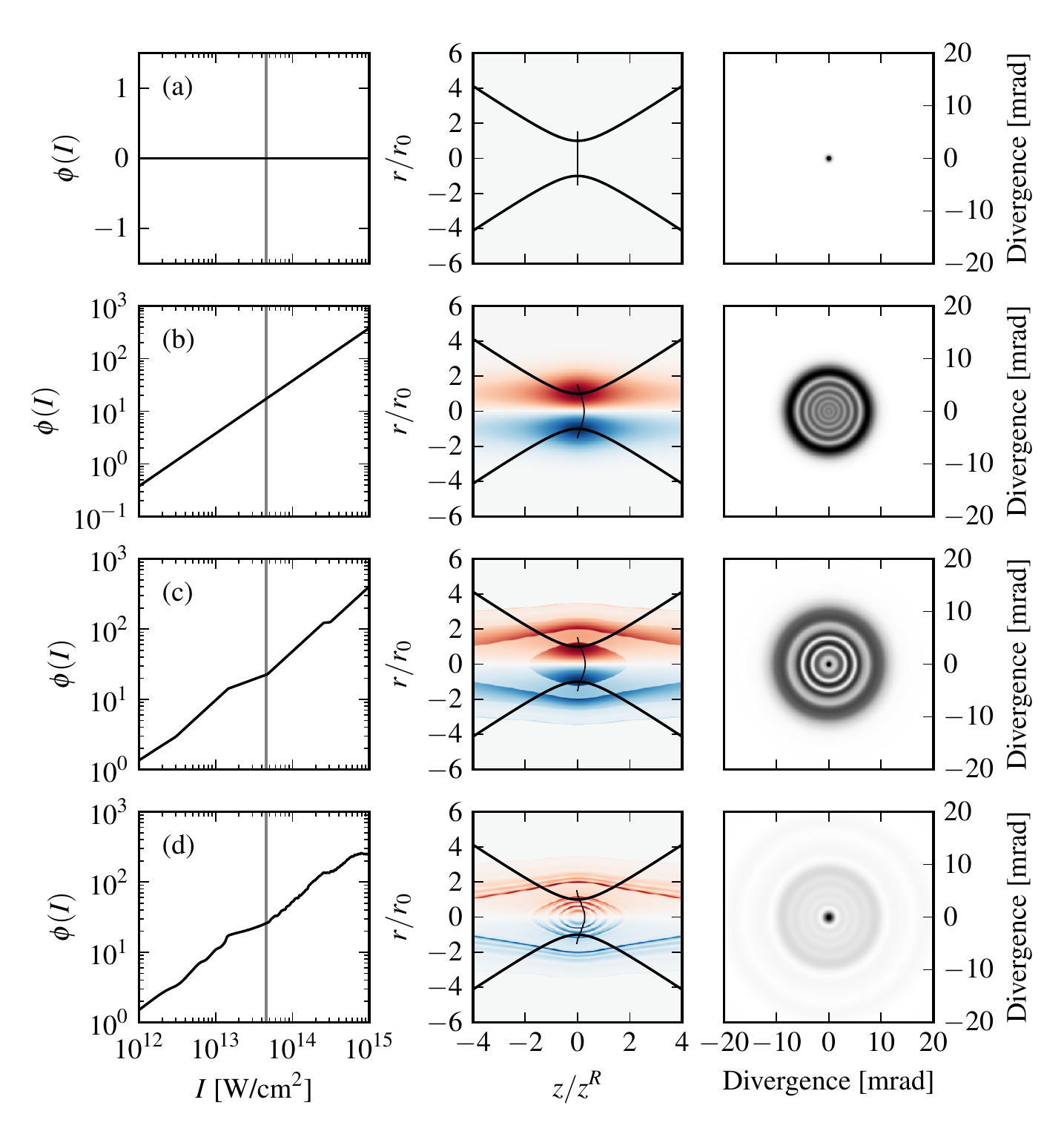}
  \caption{Explanation of how the off-axis rings come about. The
    left-most column shows the phase variation of HH25 in the focal
    plane, as a function of the fundamental intensity, for the case of
    no variation (a), a phase proportional to the intensity (b), a
    crude fit to the phase as calculated by the SFA (c) and the full
    SFA phase (d). The second case corresponds to
    \eqref{eq:DipolePhase}. The grey, vertical line indicates the
    cut-off intensity for HH25; for lower intensities, HH25 is a
    cut-off harmonic, while for higher intensities, it is in the
    plateau regime. The middle column indicates with solid
    black lines, the beam waist of the driving field as a function of
    \(z\), and the wavefront in the focal plane. The colour map behind
    shows \(\pd{\phi}{r}\), which is related to the \(k\) vector;
    emission from areas of the same colour will have the same
    direction. The right-most column shows the far-field
    amplitude. With a flat phase in the focal plane, the Gaussian
    shape will be preserved. With a simple Gaussian phase (as the
    intensity profile of the fundamental is Gaussian) in the
    near-field, ring structures will appear in the far-field
    amplitude. With more complicated phase behaviour in the
    near-field, the far-field amplitude will also be more
    complicated.}
  \label{fig:far-field-propagation}
\end{figure}

Propagating a Gaussian profile with a non-flat phase variation gives
modulation of the far-field amplitude, as seen in
figure~\ref{fig:far-field-propagation}. Depending on the form of the
near-field phase variation with the spatial profile, different
structures appear.
\clearpage

\begin{figure}[h]
  \centering
  \includegraphics{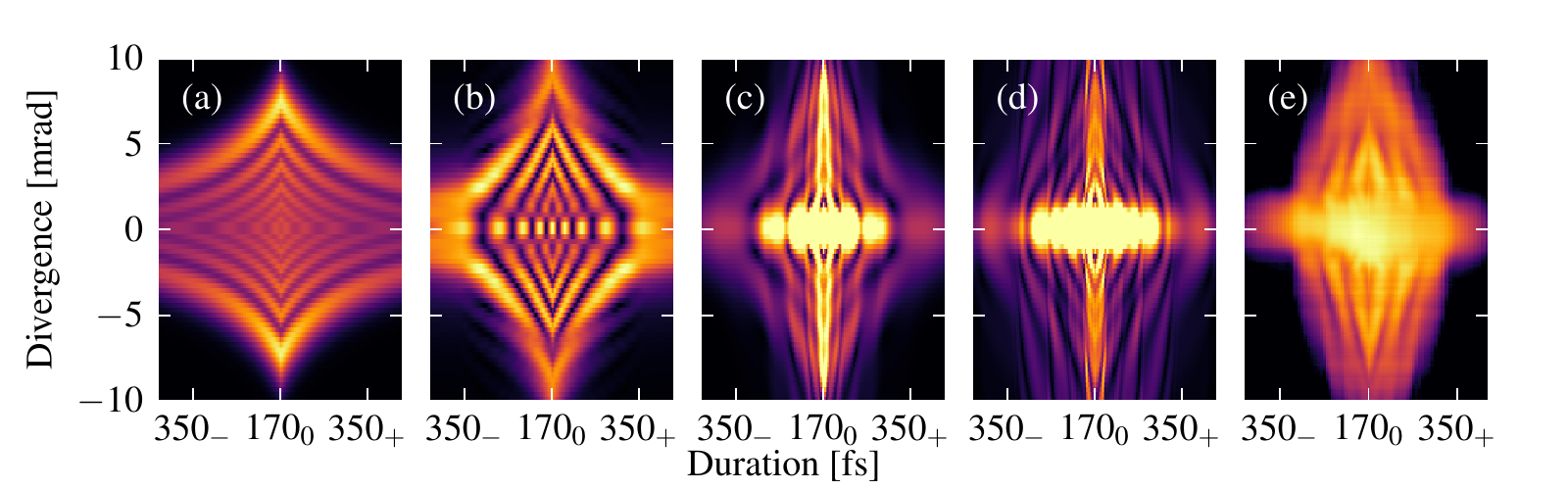}
  \caption{Far-field propagation of HH25 for a few different cases:
    (a) Gaussian model, long trajectory only
    [figure~\ref{fig:far-field-propagation}~(b)], (b) SFA log-fit
    phase [figure~\ref{fig:far-field-propagation}~(c)], (c) SFA
    [figure~\ref{fig:far-field-propagation}~(d)], (d) TDSE, (e)
    experiment. The short--long trajectory contributions has been
    saturated to focus on the off-axis interference; the colour scale
    of the theory (a--d) is linear while that of the experiment (e) is
    logarithmic as above, due to the much higher dynamic range of the
    theory.}
  \label{fig:long--long-interference}
\end{figure}

In figure~\ref{fig:long--long-interference}, the interference pattern
for \(q=25\) is displayed, from the experiment as well as calculated
using the adiabatic model, for a few different phase variations with
the spatial profile. Whereas both the TDSE and the SFA qualitatively
agree quite well with the experiment, the Gaussian beam model does
not. It is thus necessary, but not sufficient, to include higher-order
terms in the expansion of the intensity profile. Indeed, one must also
include higher-order terms in the variation of the phase with the
intensity. For the short--long interference, this mainly takes place
where two trajectories actually \emph{exist}, namely in the plateau
regime. The values of \(\alpha_{s/l}\) as presented in
tables~\ref{tab:SimulatedSpatial},~\ref{tab:SimulatedSpectra}, reflect
this by successfully reproducing the short--long interference, but not
the long--long, as is evident when comparing with the TDSE/SFA.

It cannot be said that the model presented in
figure~\ref{fig:far-field-propagation}~(c),
figure~\ref{fig:long--long-interference}~(b) only probes the long--long
interference, since it is a fit to the phase as given by the SFA;
however, as seen in figure~\ref{fig:quant-alphas}, the SFA
underestimates the short trajectory contribution compared the long
trajectory. A crude fit to the SFA phase would thus smooth out any
contribution of low amplitude such as the short trajectory one. We can
thus say that the model essentially shows the long trajectory
behaviour as is apparent from the emphasis on off-axis ring behaviour
and suppression of the short--long interference at
\SI{0}{\milli\radian}, which \emph{is} visible in the SFA calculation
shown in figure~\ref{fig:long--long-interference}~(c).

\section{Discussion}
\label{sec:discussion}
We find that our mathematical models agree well with the experimental
data in the central regions of the spatial and spectral lineouts
(figures \ref{fig:SpatialLineouts} and
\ref{fig:SpectralLineouts}). They show the robustness of the simple
model introduced in \cite{Gaarde1999}, even when it is extended to
chirped driving fields. It is a clear sign of QPI, similar to the one
described in \cite{Kim2013}, where QPI was studied using excitation by
a weak perturbation consisting of a laser pulse with controlled
delay. In analogy with that study, we can think of varying the chirp
of the driving pulse as the addition of a controlled perturbation to
the driving field.

As noted above, the pulse intensity is only dependent on the
magnitude of the chirp parameter $b$. This behaviour is clearly
reflected in the spatial lineouts, for which the peak intensity is the
parameter with largest influence; the lineouts are vertically and
horizontally symmetric. The data presented in
figure~\ref{fig:SpatialExp} are analogue to the intensity scans
presented in earlier work \cite{Zair2008, Schapper2010,Auguste2009,
  Xu2008,Jiang2014,He2015a}. However, the sign of the chirp parameter
is important in the spectral lineouts, which substantially differ for
negatively and positively chirped driving laser pulses.

The values found by comparison of our mathematical model with the
experimental data are in good agreement with the theoretical
prediction for $\alpha_l$ in the whole region of $q$ and for
$\alpha_s$ close to the cut-off region (see figure \ref{fig:Alphas}). Our
measured and calculated values of $\alpha_l$ are also in good agreement
with earlier experimental work~\cite{Benedetti2006,He2015a} and
theoretical
predictions~\cite{Mauritsson2004,Varju2005JoMO,Hostetter2010,
  Auguste2009}. In this study we have consistently extracted, in both
the spectral and spatial measurements, negative values for $\alpha_s$ for a
range of harmonics below and near the ionization threshold. While
negative values of $\alpha_s$ have been theoretically
predicted~\cite{Mauritsson2004,Varju2005JoMO}, this is to our
knowledge the first experimental measurement of negative $\alpha_s$. Also
our theoretical data as calculated by the TDSE yield negative values
of \(\alpha_s\) for the low orders, although not in perfect agreement with
the experimental data. It is worth noting that negative values of
$\alpha_s$ are a clear sign of interactions between the returning electron
wave packet and the ionic core, i.e.\ that the atomic potential cannot
be neglected in the description of the low-order short-trajectory
harmonics. If this effect can be reproduced with even higher
precision, it could lead to either a possible improvement of the
accuracy of the short-range part of the pseudo-potential used or point
towards the need for inclusion of multi-electron effects in the
description of the atom used in the calculations.

We have shown in this paper that it is possible to measure
\(\alpha_{s/l}\) for the different harmonics. To fully characterize
the temporal structure of the generated radiation, however, it is not
enough to determine the values of \(\alpha_{s/l}\) for the different
frequency components. One would also need to measure the value of
\(\Phi^0_{s/l}\) in~\eqref{eq:DipolePhase}. Using our method, we would
also be able to determine \(|\Phi^0_s-\Phi^0_l|\) to within \(2\pi\),
but not their absolute values, therefore prohibiting the full temporal
reconstruction. If one would have interference between, e.g., the long
trajectories of two neighbouring harmonics as was the case
in~\cite{Sansone2005}, one would be able to determine
\(\Phi^0_l(q)-\Phi^0_l(q+1)\), thereby enabling the full
reconstruction.

Under our experimental conditions, the harmonics \(q=11\) and \(q=13\)
correspond to energies below the ionization potential threshold $\Ip$
and are so called below-threshold harmonics. In both cases, we
observed the QPI mainly in the divergence lineouts. The experimental
observation and theoretical explanation of the QPI for below-threshold
harmonics were first made by D.~C.~Yost et al.~in 2009~\cite{Yost2009}
and $\alpha_l^*$ was expected to be \(\alpha_l^*\approx 2.5\pi-3\pi\) and
\(\alpha_s^*\approx0\).  However, in our model we found values
\(\alpha_l^*\approx2\pi\) and \(\alpha_s^*\approx-0.4\pi\); these values are in a good agreement
with later theoretical calculations \cite{Hostetter2010,He2015}.

The prominent ring structures, clearly observed for regions of large
spatial and spectral divergence, thus mainly due to the long
trajectory, are covered by our extended model. The rings appear when
higher orders than parabolic in the phase curvature are included in
the propagation of the Gaussian beams. To reproduce the detailed
structures of the rings, one has to also include higher orders in the
intensity dependence of the phase. This is of particular importance
for harmonics that have comparable contributions from the plateau and
cut-off regimes.
In \cite{Heyl2011},
similar structures were observed, interpreted as temporal Maker
fringes, e.g. an effect of phase matching between subsequent planes of
generation. The presence of this kind of phase matching in the present
work cannot be ruled out, but the qualitative agreement of our
theoretical results [figure~\ref{fig:long--long-interference}(c--d)]
with the experimental results
[figure~\ref{fig:long--long-interference}(e)] suggests that the
explanation presented here is viable.

\section{Summary}
\label{sec:summary}
In this paper, we have presented experimental data with interference
structures, observed in HHG from argon. The structures are of two
kinds; firstly due to QPI between the first two trajectories and
secondly due to long trajectory emission from atoms experiencing
different local field strengths. The former interference has been
systematically investigated by varying the chirp of the driving laser
pulses and the observed patterns are well explained by a simple
mathematical model based on a semi-classical description of HHG. By
careful comparison of the experimental observations with the model, we
are able to determine the dipole phase parameters $\alpha_s$ and
$\alpha_l$ for \(q=11-37\), which are in a good agreement with theoretical
predictions (\cite{Varju2005JoMO}), except for the short trajectory
contribution in the below-threshold harmonics and plateau regions,
where we found $\alpha_s^*$ to be negative with a value $\alpha_s^*\approx-0.4\pi$.

Furthermore, the long trajectory interference was successfully
modelled by taking into account phase curvature effects beyond the
parabolic term. It was shown that the variation of the dipole
phase parameters with respect to intensity has to be considered, to
obtain the right behaviour of the resultant interference patterns.


\ack 

This research was supported by the Swedish Foundation for Strategic
Research, the Swedish Research Council and the Knut and Alice
Wallenberg Foundation and by funding from the NSF under grant
PHY-1307083 and PHY-1403236. The quantum mechanical calculations were
performed at the Lunarc computing facility at Lund University, within
the supercomputing network of Sweden, SNIC, under the project SNIC
2015/1-386.

\appendix

\section{Derivation of Gaussian model}
\subsection{Divergence model}
\label{app:spatial-model}
Modelling the spatial profiles, we suppose that the main contribution
to the generated HHs arises around the temporal maximum of the laser
peak $I_0(\tau)$, that the driving laser field has a Gaussian profile in
the focal plane ($z=0$) characterized by the beam waist $r_0$, and
that the generated HH fields of the long and short trajectory
($E^q_s$, $E^q_l$) can be expressed [using \eqref{eq:DipolePhase}] as
\begin{equation}\label{eq:GeneratedField}
  \begin{aligned}
  E^q_{s/l}&=
  C_{s/l}\left[\sqrt{I_0(\tau)}\exp\left({-\frac{r^2}{r_0^2}}\right)\right]^n
  \ce^{\im\Phi_{s/l}}\\
  &=C_{s/l}\left[\sqrt{I_0(\tau)}\exp\left({-\frac{r^2}{r_0^2}}\right)\right]^n
  \ce^{\im\Phi^0_{s/l}}
  \exp\left[{\im\alpha_{s/l}I_0(\tau)\exp\left({-\frac{2r^2}{r_0^2}}\right)}\right],
  \end{aligned}
\end{equation}
where $C_{s/l}$ is a proportionality constant and $n$ is the
nonlinearity order of the HH conversion. By Taylor expansion to second
order in \(r\), the phase term
${\im\alpha_{s/l}I_0(\tau)\exp\left({-\frac{2r^2}{r_0^2}}\right)}$ can be
simplified to
${\im\alpha_{s/l}I_0(\tau)-\im2\alpha_{s/l}I_0(\tau)\frac{r^2}{r_0^2}}$.  In this
approximation the generated field has a Gaussian intensity profile, a
parabolic wavefront, and a phase offset. It is straightforward to
identify these with a wavefront and an intensity profile of a
\textit{shifted} Gaussian beam (GB), which has its waist position
located at $-z^f_{s/l}$:
\begin{equation}\label{eq:E_GB}
  \begin{aligned}
    E_{s/l}(r,z=0)=&E_{0s/l}\frac{w_{0s/l}}{w(z^f_{s/l})}\exp{\left[
        -\frac{r^2}{w^2(z^f_{s/l})}-
        \im k_qz^f_{s/l}-
        \im k_q\frac{r^2}{2R_{s/l}(z^f_{s/l})}+
        \im\zeta(z^f_{s/l})+
        \im\eta_{s/l}
      \right]}.
  \end{aligned}
\end{equation}
Subsequently the propagation of the generated HH can be treated as a
propagation of two GBs.  These sought-after GBs can be fully
characterized by the amplitudes $E_{0s/l}$, the distances of their
waists from the HH interaction region (plane) $z^f_{s/l}$, the
Rayleigh distances $z^R_{s/l}$, the wavevector of the generated HH
$k_q$ (corresponding to the wavelength $\lambda_q$), and the phases
$\eta_{s/l}$. $\zeta(z)=\arctan(z/z^R)$ is the Gouy phase.  For a thorough
treatment of GBs, we refer the reader to \cite{Saleh2007}. $k_q$ is
given and all other variables can be found by comparing
\eqref{eq:GeneratedField} in the parabolic approximation and
\eqref{eq:E_GB}.  From comparison of the spatial parts of the
equations, we get
\begin{equation}\label{eq:E_GBspatialpart}
  E_{0s/l}=C_{s/l}\frac{w(z^f_{s/l})}{w_{0s/l}}I_0^{\frac{n}{2}}(\tau),\qquad
  w(z^f_{s/l})=\frac{r_0}{\sqrt{n}},
\end{equation}
and from the phase parts we find
\begin{equation}\label{eq:E_GBphasepart}
  R_{s/l}^f=R(z^f_{s/l})=\frac{kr_0^2}{4\alpha_{s/l}I_0(\tau)},\qquad
  \eta_{s/l}=\Phi^0_{s/l} + \alpha_{s/l}I_0(\tau)+k_qz^f_{s/l}-\zeta(z^f_{s/l}).
\end{equation}
If we express the \eqref{eq:E_GBspatialpart} and
\eqref{eq:E_GBphasepart} using GBs
\[w_{s/l}(z)=w_{0s/l}\left[{1+{z^2}/{(z^R_{s/l})^2}}\right]^{1/2},\qquad
  R_{s/l}(z)=z\left[1+{(z^R_{s/l})^2}/{z^2}\right],\] we get a set of
two equations for two unknown variables $z^f_{s/l}$ and $z^R_{s/l}$
with solutions
\begin{equation}\label{eq:z-found}
  z_{s/l}^R=\frac{\pi \lambda_{q} (R_{s/l}^f)^2 r_0^2/n}{\lambda_{q}^2
    (R_{s/l}^f)^2+\pi r_0^4/n^2},
  \qquad
  z_{s/l}^f=\frac{\pi^2 R_{s/l}^f r_0^4/n^2}{\lambda_{q}^2 (R_{s/l}^f)^2+\pi r_0^4/n^2}.
\end{equation}
Finally, the generated HH field at the detector at distance $z$ can be
modelled as a sum of GBs representing short and long QP contribution,
\(E_{\textrm{detector}}(r,z)=E_s(r,z)+E_l(r,z)\), where
\begin{equation}\label{eq:EGBslfound}
  E_{s/l}(r,z)=E_{0s/l}\frac{w_{0s/l}}{w_{s/l}(\tilde{z}_{s/l})}
  \exp{\left[-\frac{r^2}{w_{s/l}^2(\tilde{z}_{s/l})}-
      \im k_{q}\tilde{z}_{s/l}-
      \im k_{q}\frac{r^2}{2R_{s/l}(\tilde{z}_{s/l})}+
      \im\zeta(\tilde{z}_{s/l})+
      \im\eta_{s/l}\right]},
\end{equation}
and \(\tilde{z}_{s/l}\equiv z+z_{s/l}^f\). The quantity that is
measured is proportional to
\begin{equation}\label{eq:E2detector}
|E_{\textrm{detector}}(r,z)|^2=|E_{s}(r,z)|^2+|E_{l}(r,z)|^2
+2|E_{s}(r,z)||E_{l}(r,z)|\cos[\chi(r,z)],
\end{equation}
where
\begin{equation}
\begin{aligned}
\chi(r,z)=&-\frac{k_q r^2}{2}\left[\frac{1}{R_{s}(\tilde{z}_{s})}-\frac{1}{R_{l}(\tilde{z}_{l})}\right]
+[\zeta(\tilde{z}_{s})-\zeta(\tilde{z}_{l})]-
[\zeta(z^f_{s})-\zeta(z^f_{l})]\\
&+(\Phi^0_s-\Phi^0_l)+
(\alpha_s-\alpha_l)I_0(\tau).
\end{aligned}
\end{equation}

\subsection{Spectral model}
\label{app:spectral-model}
Turning to the spectral behaviour of the harmonics, we can assume that
the main contribution is generated in the middle of the focus and that
we can neglect the spatial variation of $\Phi_{s/l}$. We describe the
short and long trajectory contributions as
\begin{equation}\label{eq:Es_l}
  E_{s/l}(t)=E_{0s/l}(t)\exp\left[\im q\omega(t)t+\im\Phi_{s/l}(t)\right],
\end{equation}
where $E_{0s/l}(t)$ is the amplitude of the generated field
approximated by
\begin{equation}\label{eq:Idriving}
  E_{0s/l}(t)= C_{s/l} I^{\frac{n}{2}}(t).
\end{equation}
$I(t)$ is the time-varying intensity in the middle of generation plane
and $\omega(t)$ is the frequency of the driving laser.  The
instantaneous frequency of the generated HH $\omega_{s/l}(t)$ is
determined as the time derivative of the phase of \eqref{eq:Es_l}:
\begin{equation}\label{eq:omega_s/l}
  \omega_{s/l}(t)=q\omega(t)+\alpha_{s/l}\frac{\partial I(t)}{\partial t}.
\end{equation}%
We suppose that the driving pulse is linearly chirped
\begin{equation}\label{eq:omega_chirped}
  \omega(t)=\omega_{0}+b(\tau)t,
\end{equation}
where $\omega_{0}$ is the central frequency of the driving laser field
and $b(\tau)$ is the chirp rate. The driving field intensity varies as
\begin{equation}\label{eq:I(t)}
  I(t)=I_{0}(\tau)\exp\left({-\frac{4 \ln 2}{\tau^2}t^2}\right),
\end{equation}
where $I_{0}(\tau)$ can be determined from \eqref{eq:Itau}. The chirp
rate $b$ is then related to the duration of the laser pulse
$\tau$ and to the duration of the Fourier transform limited pulse
$\tau\FT$ as
\begin{equation}\label{eq:b}
  b(\tau)=\pm\frac{4 \ln2}{\tau^2} \sqrt{\frac{\tau^2}{\tau\FT^2}-1}.
\end{equation}
In the presented lineouts, the negatively chirped pulses are on the
left side (negative sign in the above equation), whereas the
positively chirped pulses (positive sign) on the right side.

Together with \eqref{eq:omega_chirped} and the time derivative of
\eqref{eq:I(t)}, we find the instantaneous frequency of the generated
HH:
\begin{equation}\label{eq:omegaHH}
  \omega_{s/l}(t)=q\omega_0+qb(\tau)t-\alpha_{s/l}\frac{8 \ln 2}{\tau^2}I(t).
\end{equation}
For our simulation, the generated HH field is described as
\begin{equation}\label{eq:Efield}
  E_{s/l}(t) = C_{s/l}
  I^{\frac{n}{2}}(t)\exp\left[\im q\omega_0 t +
    \im \frac{qb(\tau)}{2}t^2
    +\im\alpha_{s/l} I(t)+\im\Phi^0_{s/l}\right].
\end{equation}
The far-field spectra of the generated HHs are computed as the Fourier transform
of the sum of the fields generated by the short and long trajectory
contributions
\begin{equation}\label{eq:FT}
  S(\omega)=\mathcal{F}[E_s(t)+E_l(t)],
\end{equation}
and the intensity in the far field is given by
\begin{equation}
  \label{eq:FTint}
  I(\omega) \propto |S(\omega)|^2.
\end{equation}

\clearpage

\bibliographystyle{iopart-num}
\bibliography{references}

\end{document}